\documentclass[11pt, a4paper, logo, copyright]{googledeepmind}

\pdftrailerid{redacted}

\makeatletter
\renewcommand\bibentry[1]{\nocite{#1}{\frenchspacing\@nameuse{BR@r@#1\@extra@b@citeb}}}
\makeatother

\usepackage{kantlipsum, lipsum}
\usepackage{dsfont}
\usepackage{gdm-colors}
\usepackage[utf8]{inputenc}   % 允许 UTF-8 输入
\usepackage{newunicodechar}   % 定义 Unicode 字符
\usepackage{amssymb}          % 提供部分符号支持
\usepackage{wasysym,marvosym}
 
\usepackage[authoryear, sort&compress, round]{natbib}

\usepackage{bbding}
\usepackage[utf8]{inputenc} % allow utf-8 input
\usepackage[T1]{fontenc}    % use 8-bit T1 fonts
\usepackage{hyperref}       % hyperlinks
\usepackage{url}            % simple URL typesetting
\usepackage{booktabs}       % professional-quality tables
\usepackage{nicefrac}       % compact symbols for 1/2, etc.
\usepackage{microtype}      % microtypography
\usepackage{amsmath}
\usepackage{graphicx}
\usepackage{multicol}
\usepackage[nameinlink]{cleveref}
\usepackage{bbm}
\usepackage{multirow}
\usepackage{soul}
\usepackage{float}
\usepackage{wrapfig}
\usepackage{blindtext}
\usepackage{tablefootnote}
\usepackage{amsfonts}
\usepackage[flushleft]{threeparttable}
\usepackage{colortbl}
\usepackage{mathtools,amssymb}
\usepackage{bm}
\usepackage{makecell}
\usepackage{caption}
\usepackage{capt-of}
\usepackage{array}
\usepackage{calc}      % 用于测量高度
\usepackage{caption}   % 用于 \captionof 命令
\usepackage{subcaption}  % 在导言区添加
\usepackage{xcolor,colortbl}
\usepackage[bottom]{footmisc}

\usepackage{xspace}

\newcommand{\eat}[1]{}
\crefformat{section}{\S#2#1#3}

\graphicspath{{figures/}}

\title{OneRec Technical Report}

\renewcommand{\today}{}

\author{\large OneRec Team}

\begin{abstract}
Recommender systems have been widely used in various large-scale user-oriented platforms for many years.  Over the past decade, recommendation technology has evolved from traditional heuristic-based rules to deep learning models, significantly improving recommendation accuracy. However, compared to the rapid changes and developments in the AI community, recommendation systems have not achieved a breakthrough in recent years. For instance, they still rely on a multi-stage cascaded architecture rather than an end-to-end approach, leading to computational fragmentation and optimization inconsistencies. 
Additionally, the cascading structure has hindered the effective application of key breakthrough technologies from the AI community in recommendation scenarios.

To address these issues, we propose \textbf{OneRec}, which reshapes the recommendation system through an end-to-end generative approach. Under this new architecture, we have achieved promising results. 
Firstly, we have enhanced the computational FLOPs of the current recommendation model by 10 $\times$ and have identified the scaling laws for recommendations within certain boundaries.
Secondly, reinforcement learning (RL) techniques, previously difficult to apply for optimizing recommendations, show significant potential in this framework. 
Lastly, through infrastructure optimizations, we have achieved 23.7\% and 28.8\% Model FLOPs Utilization (MFU) on flagship GPUs during training and inference, respectively, aligning closely with the LLM community. This architecture significantly reduces communication and storage overhead, resulting in operating expense (OPEX) that is only 10.6\% of traditional recommendation pipelines. Deployed in Kuaishou/Kuaishou Lite APP, it handles 25\% of total queries per second (QPS), enhancing overall App Stay Time by 0.54\% and 1.24\%, respectively.
Additionally, we have observed significant increases in metrics such as 7-day Lifetime (LT7), which is a crucial indicator of recommendation experience. We also provide practical lessons and insights derived from developing, optimizing, and maintaining a production-scale recommendation system with significant real-world impact.

\end{abstract}

\begin{document}
\maketitle

\vspace{2em}
\begin{figure}[h!]
    \centering
    \includegraphics[width=\textwidth]{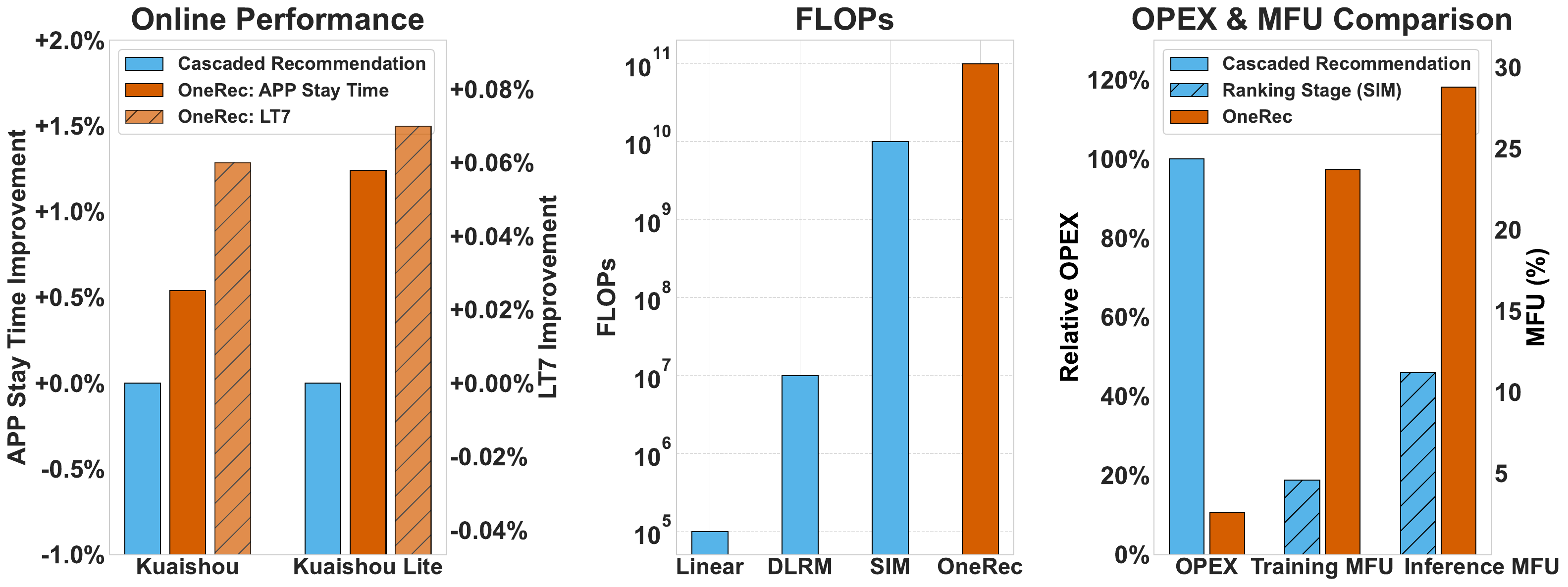}
    \captionsetup{justification=centering}
    \caption{Online performance, FLOPs, OPEX, and MFU comparison.}
    \label{fig:abs}
\end{figure}

\newpage
\setcounter{tocdepth}{2} % Set the depth of the table of contents to show sections and subsections only

\tableofcontents % Print the table of contents

% Incude paper content from external files
% \input{sections/introduction}
% \input{sections/related}
% \input{sections/training}
% %\input{sections/dataset}
% %\input{sections/method}
% \input{sections/experiment}
% % \input{sections/analysis}
% \input{sections/conclusion}
% \input{sections/template_content}
\newpage
\section{Introduction}
\begin{figure}[t]
    \centering
    \includegraphics[width=\textwidth]{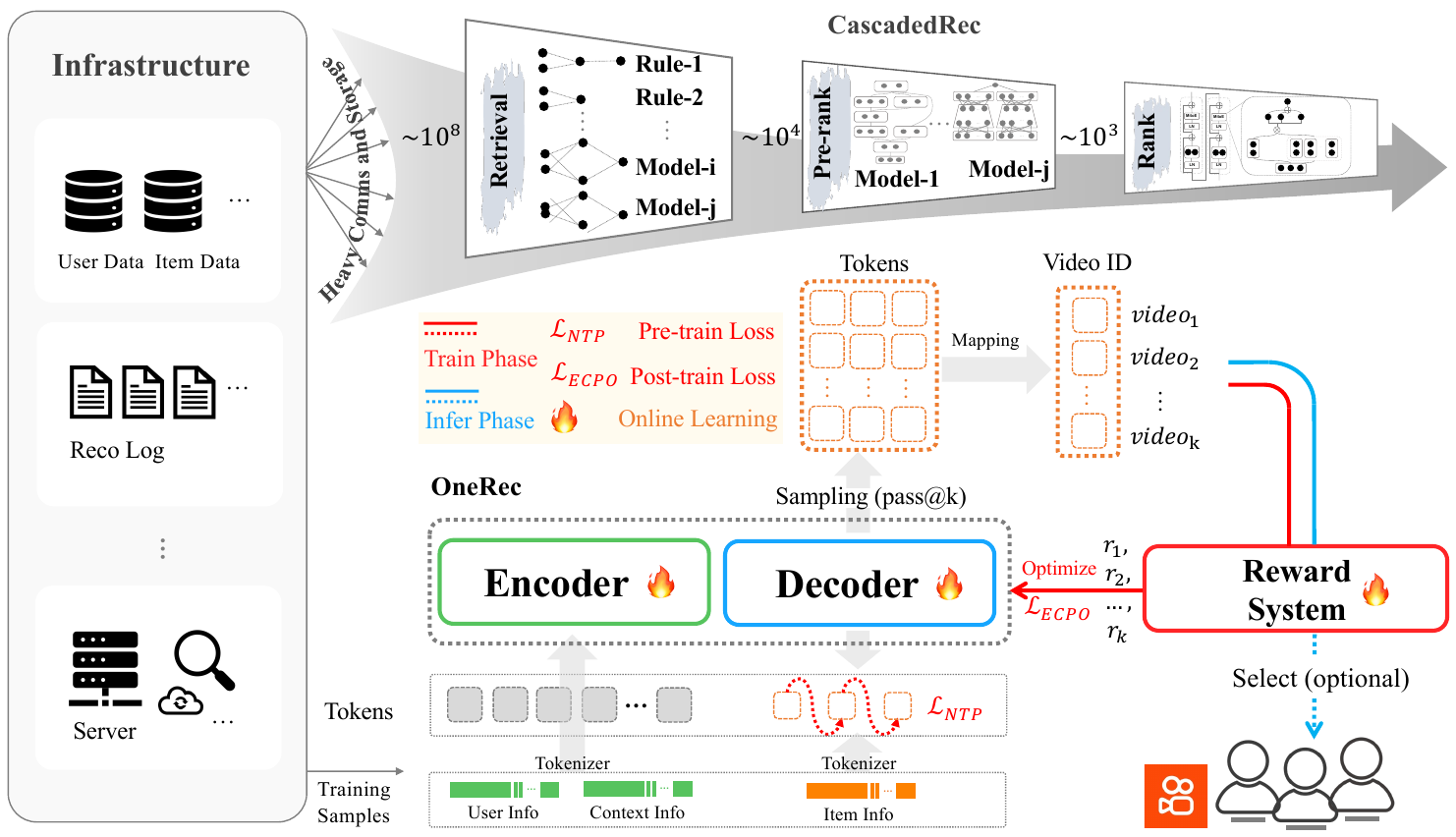}
    \caption{Comparison between a cascaded recommender system and the OneRec. The cascaded approach system typically involves stages such as retrieval, pre-ranking, and ranking, each potentially employing multiple strategies or models. In contrast, OneRec adopts an encoder-decoder architecture to generate user-preferred videos in an end-to-end manner under the guidance of a reward model.}
    \label{intro_fig}
\end{figure}

With the rapid advancement of online services, recommender systems (RS) have become essential infrastructure for mitigating information overload and delivering personalized content at scale \citep{ricci2010introduction}. 
During the past decades, recommender systems have achieved several breakthrough advancements - from early Factorization Machines \citep{rendle2010factorization} to modern deep learning architectures \citep{guo2017deepfm,cheng2016wide,zhou2018deep,pi2020search}. 
Despite the substantial progress made by the RS research community, traditional recommendation models still rely on multi-stage cascaded architectures (see the top part of Figure \ref{intro_fig}) rather than end-to-end approaches, which face several limitations that hinder their optimal performance:

\textbf{Fragmented Compute.} 
The cascaded architecture suffers from low computational efficiency. Our comprehensive analysis of resource distribution, using Kuaishou as a case study, reveals that over \textbf{50\%} of resources during serving are allocated to communication and storage rather than high-precision computation. This significant allocation to non-computational tasks highlights a fundamental inefficiency in the current architecture. Moreover, the resources dedicated to computation, particularly for the most computation-intensive ranking models, demonstrate markedly low utilization. \textbf{Specifically, the model's training and inference MFU is only 4.6\% and 11.2\% on flagship GPUs, respectively, which is substantially lower than the efficiency observed in large language models (LLMs), where the MFU is approximately 40\% on H100} \citep{shoeybi2019megatron,grattafiori2024llama}. This discrepancy underscores the inefficiency in resource utilization for computational tasks in recommender systems. Additionally, due to the high QPS requirements (\textit{greater than 400k}) and low latency demands (\textit{less than 500ms}), recommender models are often constrained to operate at a low scale and are not computation-intensive. This operational constraint further limits the potential for high-precision computation, thereby affecting the overall performance and scalability of the recommender system.

\textbf{Objective Collision.} What optimization objectives correspond to ``good'' recommendation results are not well-defined, which leads to the following conflict:
 
 1) \textit{Conflicts from Diverse Objectives}:  
 Beyond common optimization goals like click-through rate and watch time, there are competing goals (hundreds of goals in Kuaishou) from users, creators, and platform ecosystems. These objectives intervene at various stages of the system, gradually undermining system consistency and increasing complexity and operational inefficiency.

 2) \textit{Cross-Stage Modeling Conflicts}: Even when modeling similar objectives, conflicts can arise due to different structures and sizes of models at various stages. For instance, the effectiveness of the retrieval stage might be constrained by the limitations of the ranking model, which, in turn, could be affected by suboptimal upstream results. This highlights the need for a more unified optimization goal and model structure across the recommendation system to ensure coherence and efficiency.

\textbf{Lag Behind AI Evolution}. While remarkable progress has been made in LLM and visual language model (VLM) domains (e.g., scaling laws \citep{kaplan2020scaling,henighan2020scaling,hoffmann2022training}, reinforcement learning \citep{ziegler2019fine,ouyang2022training,rafailov2023direct,shao2024deepseekmath}), the existing cascaded recommendation framework presents fundamental architectural barriers to adopting these proven techniques. This structural misalignment creates a widening gap between recommendation systems and mainstream AI advancements, limiting potential performance gains from state-of-the-art approaches.

To address the challenges faced by traditional cascaded recommendation architectures, we propose \textbf{OneRec} (See the bottom part of Figure \ref{intro_fig}), a novel recommendation system designed to overcome the limitations of cascade ranking systems by integrating retrieval and ranking processes into a single-stage encoder-decoder based generative framework. This approach exhibits the following characteristics:

$\vdash$
\textbf{End-to-End Optimization}: 
The system is designed to be both end-to-end and sufficiently simple to enable direct optimization for the final objective.

$\vdash$
\textbf{Computational Efficiency}: 
With a focus on computational intensity, the method rigorously optimizes computational utilization efficiency during both training and inference phases, thereby fully leveraging the benefits brought by computing power advancements.

Our new framework yields several significant findings:  
\begin{itemize}[label=$\blacksquare$] 
\item Through extensive infrastructure optimizations, we have achieved \textbf{23.7\% and 28.8\% MFU on flagship GPUs during training and inference, respectively} --- representing 5.2$\times$ and 2.6$\times$ improvements over the original ranking model --- significantly narrowing the gap with the LLM community. More importantly, this end-to-end architecture dramatically reduces unnecessary communication and storage overhead, \textbf{resulting in OPEX that is merely 10.6\% of that associated with traditional complex recommendation pipelines}. Currently, its deployment in the main scenarios of the Kuaishou/Kuaishou Lite APP manages approximately 25\% of total QPS, \textbf{delivering improvements of 0.54\% and 1.24\% in App Stay Time}, while simultaneously improving all core metrics—including user engagement, video cold start, and distribution balance — demonstrating comprehensive performance gains.

\item We have enhanced the computational FLOPs of the current recommendation model by 10$\times$. Through this process, we have identified the scaling laws for recommendation systems. This discovery provides valuable insights into how recommendation system performance can be optimized as model size and computational resources are scaled, ensuring efficient and effective deployment in various operational contexts.

\item Reinforcement learning (RL) techniques, which previously had shown limited impact in traditional architectures, now demonstrate substantial potential within our framework. We have conducted extensive experiments with both offline and online performance comparisons and have developed specific application practices tailored to meet real-world industrial iteration requirements. These implementations enable the system to leverage RL, resulting in improved adaptability and performance.

\end{itemize}

In the remainder of this paper, we first elaborate on the OneRec architecture (Section \ref{Architecture}), detailing our tokenization pipeline for short videos, the encoder’s design for user interest modeling and compression, and scalable decoder optimization for precise output generation; we also introduce our reinforcement learning framework for recommendation optimization, discussing the impact of sampling space design, policy, and reward function on recommendation outcomes, along with empirical insights from production deployment. Next, we present the pre-training and post-training pipeline (Section \ref{Training_Framework}), covering training data construction, hyperparameter configurations, and critical implementation discussions, followed by a description of the evaluation framework (Section \ref{Evalution}), including offline metric systems and online performance/efficiency optimizations. 
Lastly, we conclude this work, discuss the existing limitations of OneRec, and propose potential
directions for future research (Section \ref{Conclusion}).

%组织结构

%	METHODS
%----------------------------------------------------------------------------------------

\section{Architecture}
\label{Architecture}
% \subsection{Overview}

In this section, we present the OneRec architecture (as illustrated in the bottom part of Figure~\ref{intro_fig}). The architecture first employs a tokenizer (Section~\ref{sec:tokenizer}) to convert videos into semantic IDs which serve as the prediction targets for the model. During the \textbf{training phase}, the encoder-decoder structure (Section~\ref{sec:enc} and Section~\ref{sec:dec}) performs next token prediction to forecast target items, while simultaneously undergoing reinforcement learning alignment through the reward system (Section~\ref{sec:reward}). In the \textbf{inference phase}, the model first generates semantic IDs and then maps these tokens back to video recommendations, with an optional reward-based selection step for further refinement.
\subsection{Tokenizer}
\label{sec:tokenizer}
OneRec is a generative recommendation system at Kuaishou, while its billion-scale, ever-growing item space prevents generating atomic identifiers due to computational and architectural constraints.
To resolve these, OneRec tokenizes items into coarse-to-fine semantic IDs using a reduced and fixed vocabulary,
enabling knowledge transfer among similar items and better generalization to new items \citep{rajput2024recommender}.
However, prior solutions \citep{rajput2024recommender,zheng2024adapting} generate semantic IDs exclusively from context features, neglecting collaborative signals and yielding suboptimal reconstruction quality, as demonstrated in Section \ref{exp:tokenizer}. Consequently, our solution integrates collaborative signals with multimodal features and then leverages RQ-Kmeans \citep{luo2024qarm} to generate higher-quality hierarchical semantic IDs.

\begin{figure}[t]
    \centering
    \includegraphics[width=1.0\textwidth]{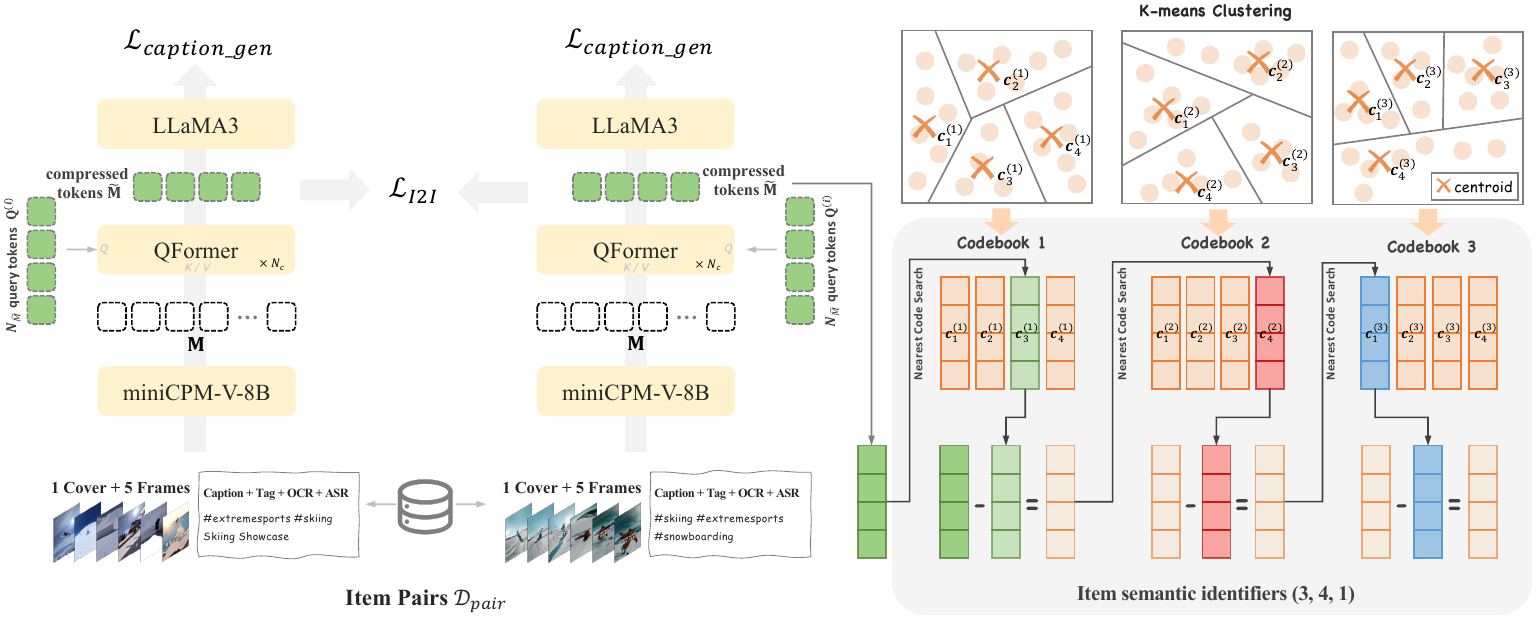}
    \caption{Illustration of our tokenizer implementation. We first align multimodal representations of item pairs with high collaborative similarity to obtain collaborative multimodal representations, then tokenize these representations into discrete semantic IDs using RQ-Kmeans.}
    \label{fig:tokenizer}
\end{figure}

\subsubsection{Aligned Collaborative-Aware Multimodal Representation}
\label{sec:ia-emb}
We integrate multimodal content with collaborative signals by aligning multimodal representations of collaboratively similar item pairs, as shown in Figure~\ref{fig:tokenizer} (left). Therefore, we require the preparation of multimodal representations, item pairs, and an alignment strategy:

\begin{itemize}
    \item \textbf{Multimodal Representations.} We incorporate multimodal inputs for each video: the caption, tag, ASR (speech-to-text), OCR (image-to-text), the cover image, and 5 uniformly sampled frames. These inputs are processed using miniCPM-V-8B \citep{hu2024minicpm}, generating $N_M=1280$ token vectors $\mathbf{M}\in\mathbb{R}^{N_M\times d_{t}}$ ($d_{t}=512$). A Querying Transformer (QFormer) \citep{li2023blip} then compresses these tokens with $N_{\tilde{M}}=4$ learnable query tokens $\mathbf{Q}^{(1)}\in\mathbb{R}^{N_{\tilde{M}}\times d_{t}}$:
        \begin{align}
           \mathbf{Q}^{(i+1)} &= \text{CrossAttn}\left(\mathbf{Q}^{(i)}, \mathbf{M},\mathbf{M}\right), \\
           \mathbf{Q}^{(i+1)} &= \text{FFN}(\text{RMSNorm}(\mathbf{Q}^{(i+1)}),
           \quad \text{for } i \in \{1, 2,\ldots, N_{c}\},
        \end{align}
    where $\tilde{\mathbf{M}}=\mathbf{Q}^{(N_{c}+1)}\in\mathbb{R}^{N_{\tilde{M}}\times d_{t}}$ denotes the compressed version of $\mathbf{M}$, and $N_{c}=4$ denotes the number of QFormer layers.
    \item \textbf{Item Pairs.} 
    We construct high-quality item-pair dataset $\mathcal{D}_{pair}$ via: 
    1) User-to-Item Retrieval: For each user, we take a positively clicked target item and pair it with the most collaboratively similar item from the user’s latest historical positive clicks, and 
    2) Item-to-Item Retrieval: We pair items exhibiting high similarity scores (e.g., the Swing similarity) \citep{DBLP:journals/corr/abs-2010-05525}.

    \item \textbf{Item-to-Item Loss and Caption Loss.} We introduce dual training objectives: 1) An item-to-item contrastive loss aligns representations of collaboratively similar video pairs $(i, j)\in\mathcal{D}_{pair}$, capturing behavioral patterns, and 2) a caption loss prevents hallucination by performing next-token prediction on video captions using LLaMA3 \citep{dubey2024llama} as the decoder, thereby preserving content understanding capabilities.
    \begin{equation}
        \mathcal{L}_{I2I} = -\frac{1}{\left|\mathcal{B}\right|}
                                    \sum_{(i,j)\in\mathcal{B}}
                                    \log
                                    \frac{
                                        \exp
                                        \left(
                                            \text{sim}\left(
                                                \tilde{\mathbf{M}}_i, \tilde{\mathbf{M}}_j
                                            \right)/\tau
                                        \right)
                                        }
                                        {
                                            \sum_{(i',j')\in\mathcal{B}}
                                            \exp
                                            \left(
                                                \text{sim}\left(
                                                    \tilde{\mathbf{M}}_{i}, \tilde{\mathbf{M}}_{j'}
                                                \right)/\tau
                                            \right)
                                        },
    \end{equation}
    \begin{equation}
        \mathcal{L}_{caption\_gen} = -\sum_{k} \log P\left(t^{k+1}|\left[t^1, t^2, \cdots, t^k\right]\right),
    \end{equation}
    where $\tau$ denotes the temperature coefficient, $\text{sim}(\cdot, \cdot)$ denotes the similarity function, $\mathcal{B}$ denotes a batch of $\mathcal{D}_{pair}$, $t^k$ denotes the $k$-th caption token.
\end{itemize}

\subsubsection{Tokenization}
\label{sec:tokenization}
We utilize RQ-Kmeans \citep{luo2024qarm} for tokenization, which employs residual quantization to generate semantic IDs in a coarse-to-fine manner. This method constructs codebooks by applying K-means clustering directly on the residuals. An illustration of the RQ-Kmeans process is provided in Figure~\ref{fig:tokenizer} (right).

Formally, the initial residual at layer $l=1$ is defined as:
\begin{equation}
    \mathcal{R}^{(1)} = \left\{ \tilde{\mathbf{M}}_i \in \mathbb{R}^{N_{\tilde{M}} \times d_{t}} \mid \forall \text{ video }i \right\}.
\end{equation}
For each layer $l$, the codebook $\mathcal{C}^{(l)}$ is derived from K-means centroids of $\mathcal{R}^{(l)}$:
\begin{equation}
    \mathcal{C}^{(l)} = \text{K-means}\left( \mathcal{R}^{(l)}, N_t \right),
\end{equation}
where $\mathcal{C}^{(l)} = \left\{ \bm{c}^{(l)}_k \in \mathbb{R}^{N_{\tilde{M}}\times d_{t}} \mid k = 1, \dots, N_t \right\}$ and $N_t$ is the codebook size. The nearest centroid index for item $i$ is computed as:
\begin{equation}
    s^{l}_i = \arg\min_{k} \left\| \mathcal{R}^{(l)}_i - \bm{c}^{(l)}_k \right\|,
\end{equation}
where $\| \cdot \|$ denotes the Euclidean norm. The residual of video $i$ for layer $l+1$ is then updated:
\begin{equation}
    \mathcal{R}^{(l+1)}_i = \mathcal{R}^{(l)}_i - \bm{c}^{(l)}_{s^{l}_i}.
\end{equation}
This quantization iterates across $L_t=3$ layers.

As demonstrated in Section \ref{exp:tokenizer}, RQ-Kmeans offers enhanced reconstruction quality, better codebook utilization, and improved balance compared to the widely used RQ-VAE \citep{rajput2024recommender, lee2022autoregressive}. At this stage, each video $m$ can be represented by $L_t$ coarse-to-fine semantic identifiers: $\{s^{1}_m, s^{2}_m, \ldots, s^{L_t}_m\}$, which will serve as the output of the OneRec recommendation system, enabling progressive item generation. 

% encoder-decoder
\begin{figure}[t]
    \centering
    \includegraphics[width=\textwidth]{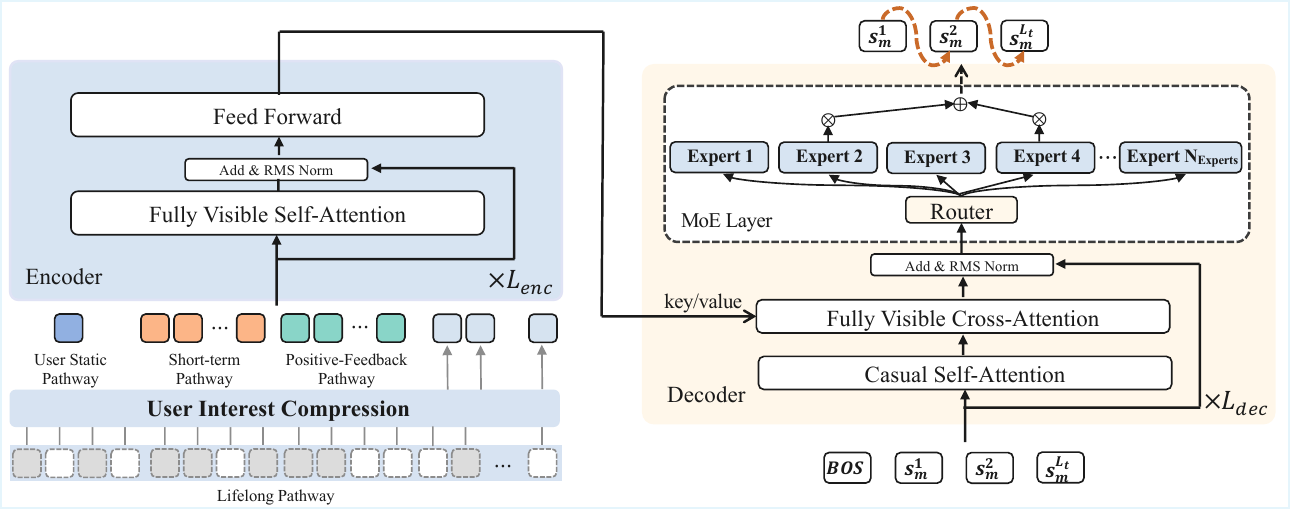}
    \captionsetup{justification=centering}
    \caption{Illustration of our encoder-decoder architecture.}
    \label{fig:encdec}
\end{figure}

\subsection{Encoder}
\label{sec:enc}

\subsubsection{Multi-Scale Feature Engineering} \label{multiscale_feature}

This section presents the feature engineering component of OneRec. We process user behavior data through four specialized embedding pathways, each designed to capture distinct scales of user interaction patterns: \textbf{user static pathway, short-term pathway, positive-feedback pathway, and lifelong pathway}.

\paragraph{User Static Pathway} 
The user static pathway generates a compact representation of core user characteristics, incorporating user identifier (\texttt{uid}), age (\texttt{age}), gender (\texttt{gender}), etc., which is then transformed into the model's hidden dimension:
\begin{align}
\mathbf{f}_u &= [\mathbf{e}_{\text{uid}}; \mathbf{e}_{\text{gender}}; \mathbf{e}_{\text{age}};\cdots], \\
\mathbf{h}_u &= \text{Dense}(\text{LeakyReLU}(\text{Dense}(\mathbf{f}_u))).
\end{align}
where $\mathbf{e}_{\text{uid}},\mathbf{e}_{\text{gender}}, \mathbf{e}_{\text{age}} \in \mathbb{R}^{64}$ and $\mathbf{h}_{u} \in \mathbb{R}^{1 \times d_{\text{model}}}$.

\paragraph{Short-term Pathway} 
The short-term behavior pathway processes the most recent ($L_{s}=20$) user interactions, incorporating video identifier (which can be represented as video identifiers \texttt{vid} or semantic identifiers \texttt{sid} as described in Section~\ref{sec:tokenization}, we will discuss these two representation approaches in Section~\ref{sec:sid}.), author identifiers (\texttt{aid}), tags (\texttt{tag}), timestamps (\texttt{ts}), playtime (\texttt{playtime}), duration (\texttt{dur}), labels (\texttt{label}, user interactions with each video, including like, follow, forward, dislike, comment,  profile entry, etc.) This pathway produces representations that capture immediate user preferences and contextual factors influencing current behavior patterns:
\begin{align}
\mathbf{f}_s &= [\mathbf{e}_{\text{vid}}^s; \mathbf{e}_{\text{aid}}^s; \mathbf{e}_{\text{tag}}^s; \mathbf{e}_{\text{ts}}^s; \mathbf{e}_{\text{playtime}}^s;\mathbf{e}_{\text{dur}}^s; \mathbf{e}_{\text{label}}^s], \\
\mathbf{h}_s &= \text{Dense}(\text{LeakyReLU}(\text{Dense}(\mathbf{f}_s))),
\end{align}
The feature dimensions are organized as follows: video embeddings $\mathbf{e}_{\text{vid}}^s$ match the model dimension $d_{\text{model}}$, author embeddings $\mathbf{e}_{\text{aid}}^s$ use 512 dimensions, while all remaining features employ 128 dimensions. All features span $L_s$ sequence positions, yielding the final representation $\mathbf{h}_{s} \in \mathbb{R}^{L_s \times d_{\text{model}}}$.

\paragraph{Positive-feedback Pathway} 
The positive-feedback behavior pathway operates on a sequence of high-engagement interactions ($L_p = 256$). The pathway maintains the established dimensional structure:
\begin{align}
\mathbf{f}_p &= [\mathbf{e}_{\text{vid}}^p; \mathbf{e}_{\text{aid}}^p; \mathbf{e}_{\text{tag}}^p; \mathbf{e}_{\text{ts}}^p; \mathbf{e}_{\text{playtime}}^p;\mathbf{e}_{\text{dur}}^p; \mathbf{e}_{\text{label}}^p], \\
\mathbf{h}_p &= \text{Dense}(\text{LeakyReLU}(\text{Dense}(\mathbf{f}_p))).
\end{align}
All features span $L_p$ sequence positions, yielding the final representation $\mathbf{h}_{p} \in \mathbb{R}^{L_p \times d_{\text{model}}}$.

\paragraph{Lifelong Pathway} 
The lifelong behavior pathway is designed to process ultra-long user interaction histories with sequences of up to 100,000 videos. Directly applying attention mechanisms to such sequences is computationally prohibitive. This pathway employs a two-stage hierarchical compression strategy inspired by our previous work \citep{si2024twin}.

\textbf{Behavior Compression } Using the multimodal content representations described in Section~\ref{sec:ia-emb}, we perform hierarchical K-means clustering on each user's interaction sequence. To balance computational efficiency and model effectiveness, we dynamically adjust the number of clusters by setting the cluster count for each step to $\lfloor\sqrt[3]{|D|} \rfloor$, where $|D|$ is the number of items in the current data. This is an empirically determined setting. The clustering process terminates when the number of items in the current cluster does not exceed a preset threshold $M$. Upon termination, we select the item closest to each cluster center as the representative of that cluster.

\textbf{Feature Aggregation }
For each cluster, we construct representative features by handling discrete and continuous attributes differently. For sparse categorical features such as \texttt{vid}, \texttt{aid}, and \texttt{label}, we directly inherit the features from the representative video (i.e., the video closest to the cluster center). For continuous features such as \texttt{tag}, \texttt{ts}, \texttt{playtime}, and \texttt{duration}, we compute the average values across all videos within the cluster to capture collective behavioral patterns. 

For the user's long-term historical sequence ($L_l = 2000$), each video is replaced by the features of its corresponding cluster representative:
\begin{align}
\mathbf{f}_l &= [\mathbf{e}_{\text{vid}}^l; \mathbf{e}_{\text{aid}}^l; \mathbf{e}_{\text{tag}}^l; \mathbf{e}_{\text{ts}}^l; \mathbf{e}_{\text{playtime}}^l;\mathbf{e}_{\text{dur}}^l; \mathbf{e}_{\text{label}}^l], \\
\mathbf{v}_l &= \text{Dense}(\text{LeakyReLU}(\text{Dense}(\mathbf{f}_l))).
\end{align}
The final representation $\mathbf{v}_{l} \in \mathbb{R}^{L_l \times d_{\text{model}}}$.
The lifelong pathway compresses historical sequences through QFormer, where learnable query vectors $\mathbf{h}_l^{(0)} \in \mathbb{R}^{N_q \times d_{\text{model}}}$ ($N_q=128$) attend to the processed historical features:
\begin{align}
\mathbf{h}_l^{(i+1)} &= \text{CrossAttn}(\mathbf{h}_l^{(i)}, \mathbf{v}_l, \mathbf{v}_l), \\
\mathbf{h}_l^{(i+1)} &= \text{FFN}(\text{RMSNorm}(\mathbf{h}_l^{(i+1)})).
\end{align}
Followed by $N_l = 2$ blocks, we obtain the compressed lifelong feature representation $\mathbf{h}_l = \mathbf{h}_l^{(N_l)} \in \mathbb{R}^{N_q \times d_{\text{model}}}$.

\subsubsection{Encoder Architecture}
As illustrated in Figure~\ref{fig:encdec}, the encoder architecture of OneRec integrates multi-scale user behavior representations through a unified transformer-based framework. The encoder concatenates the outputs from the four multi-scale pathways to form a comprehensive input sequence:
\begin{equation}
\mathbf{z}^{(1)} = [\mathbf{h}_{u}; \mathbf{h}_s; \mathbf{h}_p; \mathbf{h}_l] + \mathbf{e}_{\text{pos}}
\end{equation}
where $\mathbf{e}_{\text{pos}} \in \mathbb{R}^{(1 + L_s + L_p + N_q) \times d_{\text{model}}}$ represents learnable positional embeddings. The integrated representation is processed through $L_{\text{enc}}$ transformer encoder layers, each consisting of fully visible self-attention mechanisms followed by feed-forward networks with RMS normalization:
\begin{align}
\mathbf{z}^{(i+1)} &= \mathbf{z}^{(i)} + \text{SelfAttn}(\text{RMSNorm}(\mathbf{z}^{(i)})),\\
\mathbf{z}^{(i+1)} &= \mathbf{z}^{(i+1)} + \text{FFN}(\text{RMSNorm}(\mathbf{z}^{(i+1)})).
\end{align}
The final encoder output $\mathbf{z}_{\text{enc}}=\mathbf{z}^{(L_{enc}+1)} \in \mathbb{R}^{(1 + L_s + L_p + N_q) \times d_{\text{model}}}$ provides a holistic multi-scale user behavior representation, serving as the foundation for subsequent recommendation generation.

\subsection{Decoder}
\label{sec:dec}

OneRec adopts a point-wise generation paradigm during the decoding phase. For each target video $m$, the decoder input sequence is constructed by concatenating a learnable beginning-of-sequence token with the video's semantic identifiers:
\begin{align}
    \mathcal{S}_m &= \left\{s_{[\mathrm{BOS}]}, s^1_m, s^2_m, \cdots, s^{L_t}_m\right\}, \\
    \mathbf{d}^{(0)}_m &= \text{Emb\_lookup}(\mathcal{S}_m).
\end{align}
The decoder processes this sequence through $L_{\text{dec}}$ transformer layers. Each layer performs sequential operations:
\begin{align}
\mathbf{d}^{(i+1)}_m &= \mathbf{d}^{(i)}_m + \text{CausalSelfAttn}(\mathbf{d}^{(i)}_m), \\
\mathbf{d}^{(i+1)}_m &= \mathbf{d}^{(i+1)}_m + \text{CrossAttn}(\mathbf{d}^{(i+1)}_m, \mathbf{Z}_{\text{enc}}, \mathbf{Z}_{\text{enc}}), \\
\mathbf{d}^{(i+1)}_m &= \mathbf{d}^{(i+1)}_m + \text{MoE}(\text{RMSNorm}(\mathbf{d}^{(i+1)}_m)).
\end{align}
Each decoder layer incorporates a Mixture of Experts (MoE) feed-forward network to enhance model capacity while maintaining computational efficiency. The MoE layer employs $N_{\text{experts}}$ expert networks with a top-$k$ routing strategy:
\begin{align}
\text{MoE}(\mathbf{x}) = \sum_{j=1}^{k} \text{Gate}_j(\mathbf{x}) \cdot \text{Expert}_j(\mathbf{x}),
\end{align}
where $\text{Gate}_j(\mathbf{x})$ represents the gating weights determined by the routing mechanism, and $\text{Expert}_j(\mathbf{x})$ denotes the output of the $j$-th selected expert network. To ensure balanced expert utilization without introducing interference gradients, we implement a loss-free load balancing strategy following \citep{liu2024deepseek}.

The model is trained using cross-entropy loss for next-token prediction on the semantic identifiers of target video $m$:
\begin{equation}
\mathcal{L}_{\mathrm{NTP}}=- \sum_{j=0}^{L_t-1} \log P\left(s_m^{j+1} \mid\left[s_{[\mathrm{BOS}]}, s^1_m, s^2_m, \cdots, s^j_m \right] \right)
\label{eq:ntp}
\end{equation}

\subsection{Reward System}
\label{sec:reward}
The pre-trained model only fits the distribution of the exposed item space through next token prediction, and the exposed items are obtained from the past traditional recommendation system. This results in the model being unable to break through the ceiling of traditional recommendations. To address this issue, we introduce preference alignment based on a reward system, using on-policy reinforcement learning to train the model in the generated item space. Through rewards, the model perceives more fine-grained preference information. We introduce the preference reward to align user preferences, the format reward to ensure the generation format is as legal as possible, and the specific industrial reward to align with some special industrial scenario needs.
\begin{figure}[t]
\centering
\includegraphics[width=\textwidth]{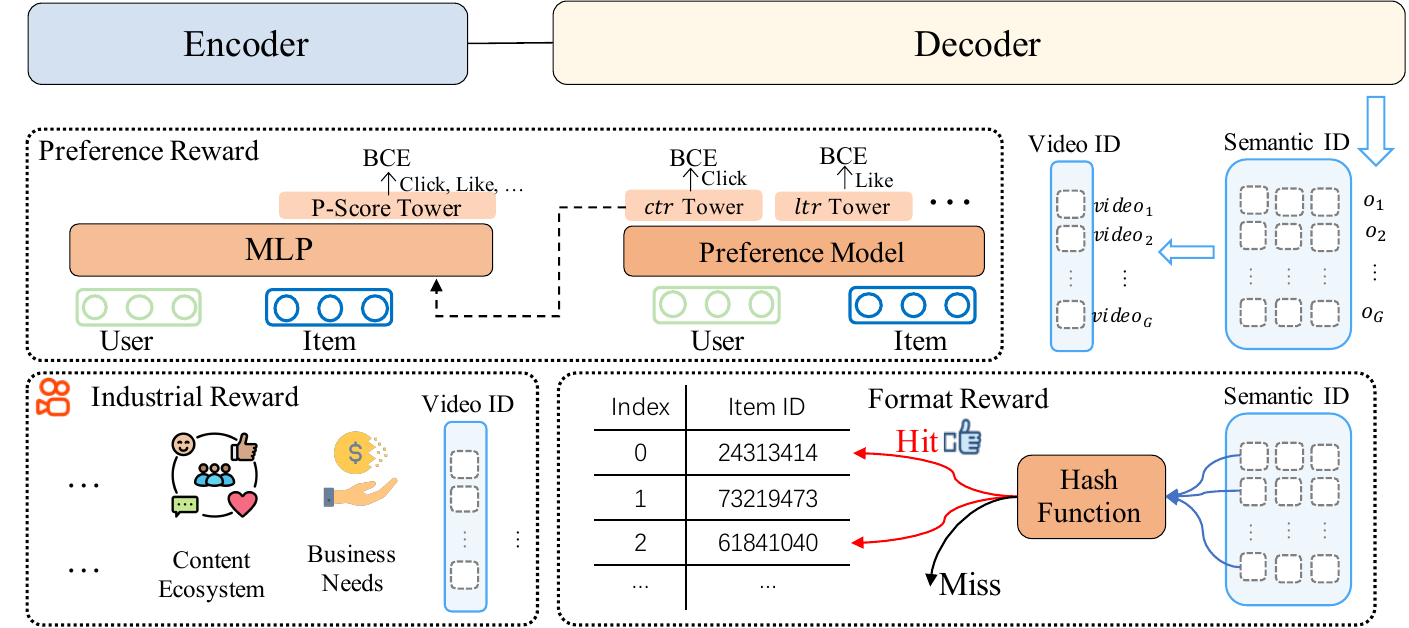}
\caption{Overall Framework of the Reward System. The Reward System is composed of three parts. They assign Preference Reward (P-Score), Format Reward, and specific Industrial Reward to the videos generated by the model, respectively.}
\label{fig:reward-system}
\end{figure}

\subsubsection{User Preference Alignment} \label{fusion_score}
In recommendation systems, defining a "good recommendation" is much more challenging than determining the correctness of a mathematical solution. Traditional approaches ~\citep{wang2024home, chang2023twin} often define multiple objectives, such as clicks, likes, comments, and watch time, which are then combined into a score through a weighted fusion of the predicted values (xtr) for each objective. However, manually tuning these fusion weights is challenging, not only lacking accuracy but also lacking personalization, and often results in optimization conflicts between objectives.

To address these limitations, we propose using a neural network to learn a personalized fusion score, referred to as P-Score (Preference Score)~\citep{cao2025pantheon}. The overall framework of this model is illustrated in Figure \ref{fig:reward-system} (middle). The model's underlying architecture is based on the Search-based Interest Model (SIM) ~\citep{pi2020search}. It includes multiple towers, each dedicated to learning specific objectives. During training, these towers compute binary cross-entropy (BCE) loss using the corresponding objective labels as auxiliary tasks. The hidden states of each tower, along with user and item representations, are fed into the final layer's Multi-Layer Perceptron (MLP). This MLP is followed by a single tower outputting the P-Score, which computes binary cross-entropy loss using the labels of all objectives.The loss can be formally represented as follows: 
\begin{equation}
\mathcal{L}_{\mathrm{P\text{-}Score}} =\sum_{xtr\in S_o}w^{\mathrm{xtr}}\mathcal{L}_{\mathrm{P\text{-}Score}}^{\mathrm{xtr}}
\end{equation}
\begin{equation}
\mathcal{L}_{\mathrm{P\text{-}Score}}^{\mathrm{xtr}} =-(y^{\mathrm{xtr}}\log{p}+(1-y^{\mathrm{xtr}})\log{(1-p)}), \\  
\end{equation}
\begin{equation}
S_o = \{\mathrm{ctr,lvtr,ltr,vtr,...\}} 
\end{equation}
We adjust the value of $w^{\mathrm{xtr}}$ to bias the P-Score towards each objective, ultimately achieving an improvement in AUC across all objectives.
This method allows the model to receive specific user information and adjust the Preference Score for that user appropriately, without compromising the experience of other users. Compared to the previous approach of indiscriminate weighted summation, this method is more likely to achieve Pareto optimization. Therefore, we use the P-Score obtained by this method as the reward for preference alignment.

\paragraph{Early Clipped GRPO}
In this section, we introduce how to use the Preference Score to align user preferences. We use ECPO (\underline{E}arly \underline{C}lipped GR\underline{PO}) for optimization. Specifically, for a user $u$, we generate $G$ items using the old policy model. Each item, along with the user, is input into the Preference Reward Model to obtain the P-Score as reward $r_i$.  The optimization objective is as follows:
\begin{equation}
\label{eq:ecpo}
\mathcal{J}_{ECPO}(\theta) = \mathbb{E}_{u\sim P(U),\{o_i\}_{i=1}^G\sim\pi_{\theta_{old}}}\left[\frac1G\sum_{i=1}^G\text{min}\left(\frac{\pi_{\theta}(o_i|u)}{\pi_{\theta_{old}}'(o_i|u)}A_i,\text{clip}\left(\frac{\pi_{\theta}(o_i|u)}{\pi_{\theta_{old}}'(o_i|u)},1-\epsilon,1+\epsilon\right)A_i\right)\right],    
\end{equation}

\begin{equation}
    A_i = \frac{r_i-\text{mean}(\{r_1,r_2,...,r_G\})}{\text{std}(\{r_1,r_2,...,r_G\})},
\end{equation}

\begin{equation}
    \pi_{\theta_{old}}'(o_i|u)=\text{max}\left(\frac{\text{sg}(\pi_\theta(o_i|u))}{1+\epsilon+\delta},\pi_{\theta_{old}}(o_i|u)\right), \quad\delta>0,
\end{equation}
where sg represents the stop gradient operation and $\delta$ is a hyperparameter greater than 0.

\begin{figure}[t]
\centering
\includegraphics[width=0.8\textwidth]{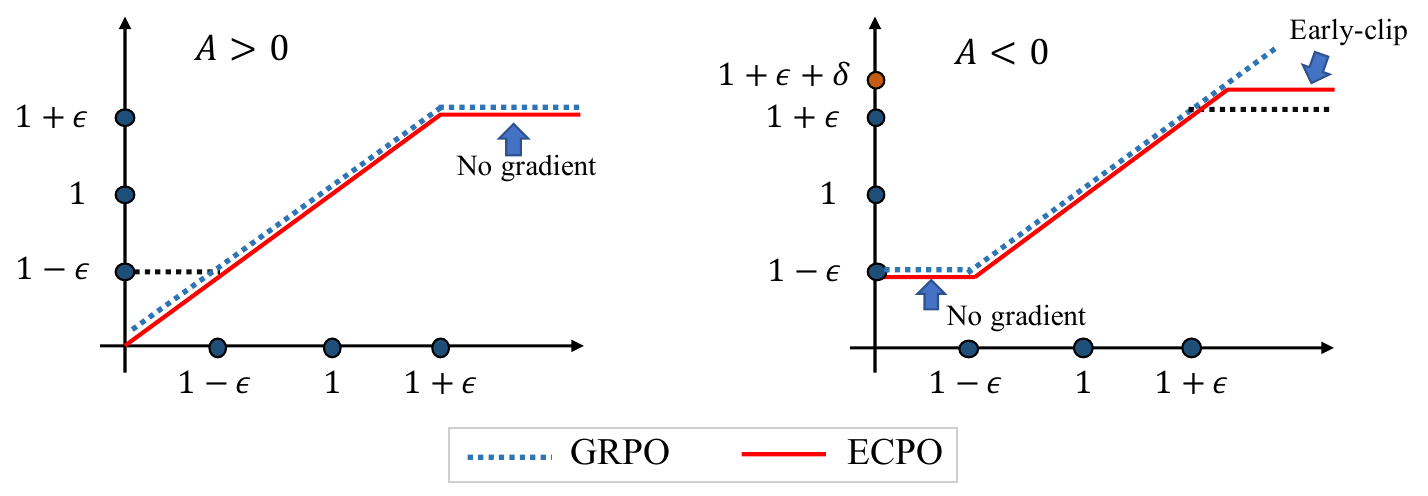}
\caption{Illustration of ECPO. The $x$-axis is $\pi_{\theta} / \pi_{\theta_{old}}$ and the $y$-axis is the clipped $\pi_{\theta} / \pi_{\theta_{old}}$. Items with \( A > 0 \) are processed in the same way as the original GRPO, while items with \( A < 0 \) are constrained by early-clipping to limit the maximum ratio.}
\label{fig:ecpo}
\end{figure}

We make a modification to GRPO (Group Policy Relative Optimization) \citep{liu2024deepseek} to make its training process more stable. The illustration is presented in Figure \ref{fig:ecpo}. In the original GRPO, a large policy ratio ($\pi_{\theta}/\pi_{\theta_{old}}$) is allowed for negative advantages, which can easily lead to gradient explosion. Therefore, we preemptively clip policies with large ratios to ensure training stability while still allowing corresponding negative advantages to take effect. The larger the $\delta$, the larger the tolerable policy ratio, which means the larger the tolerable gradient. This can be determined based on actual needs. In OneRec, we set $\delta$ to 0.1, which indicates that the ratio of policies with negative advantages is allowed to slightly exceed $1+\epsilon$. We remove the KL divergence loss because the Reinforcement Learning (RL) and Supervised Fine-Tuning (SFT) are trained together in OneRec, and the SFT loss ensures the model remains stable.

\subsubsection{Generation Format Regularization}
\label{sec:format}
In generative recommendation, the legality ratio refers to the proportion of generated semantic ID sequences that can be mapped to actual item IDs. This metric is crucial for assessing the stability of generation. 
In practice, the cardinality of semantic ID sequences $N_t^{L_t}$ is much larger than that of videos.
This ensures that all items are covered, and a larger vocabulary introduces more parameters, leading to better performance. However, this may also result in generating semantic ID sequences without corresponding item IDs during inference, i.e., illegal generation. 

\begin{figure}[h]
\centering
\includegraphics[width=\textwidth]{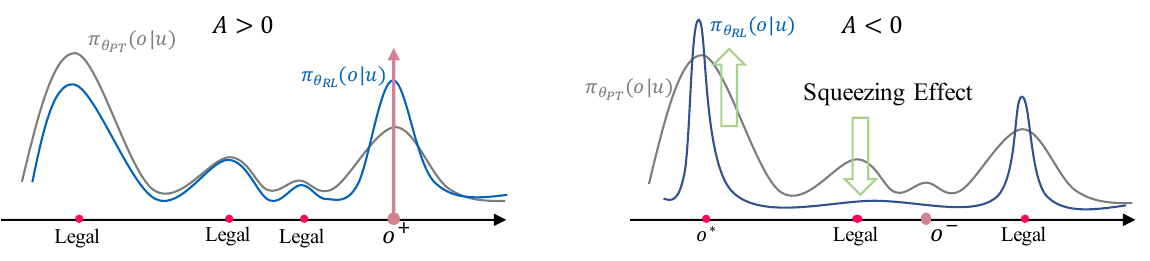}
\caption{Illustration of squeezing effect. $\pi_{\theta_{PT}}$ represents the pre-trained model, while $\pi_{\theta_{RL}}$ represents the model trained with ECPO. $o^+$ refers to videos with positive advantages, while $o^-$ refers to those with negative advantages.}
\label{fig:squeeze}
\end{figure}

Introducing reinforcement learning with ECPO significantly increases the generation of illegal outputs. Recent work \citep{ren2024learning} suggests that this is due to the squeezing effect caused by negative advantages. As shown in Figure \ref{fig:squeeze}, the pre-trained model has learned to generate most of the legal tokens. After incorporating RL, items with \( A > 0 \) only slightly adjust the distribution. When an item with \( A < 0 \) is applied, the model's probability distribution compresses most of the probability mass into what it currently considers the optimal output $o^*$. This results in the probabilities of some legal tokens being squeezed to levels comparable to those of illegal tokens, making it difficult for the model to distinguish legal tokens. 

To address this issue, we propose incorporating a format reward in reinforcement learning to encourage the model's legal generation. Specifically, we randomly select $K$ samples from the $G$ samples for legality reinforcement learning. For legal samples, we set the advantage to 1, and for illegal samples, we discard them directly to avoid the squeezing effect.
\begin{equation}
    A_i =
\begin{cases} 1 & \text{if } o_i\in I_\text{legal} \\
0 & \text{if } o_i\notin I_\text{legal}
\end{cases}
\end{equation}
The optimization objective formulation is the same as the ECPO (Equation~\ref{eq:ecpo}) and we directly use $A_i$ as advantages.
\subsubsection{Industrial Scenario Alignment}

In industrial scenarios, the recommendation system needs to consider not only user preferences but also various other aspects. For example, at Kuaishou, the ecosystem of the video community, commercialization needs, and the delivery of cold-start and long-tail videos. Traditional recommendation systems attempt to address these issues by applying algorithms or strategies at one stage of the recommendation pipeline. Due to inconsistencies across different stages, this can easily lead to a recurring cycle of unexpected problems emerging alternately. Engineers are forced to constantly make adjustments through patching, resulting in a bloated system over time that hinders iteration. In OneRec, we only need to incorporate optimization objectives into the reward system and adopt reinforcement learning to perform targeted optimization. This approach is not only convenient but also allows for end-to-end implementation, maintaining system consistency. We will provide an example of optimization practice in Section \ref{sec:sir_exp}.

\section{Training Framework}
\label{Training_Framework}
\subsection{Training Infrastructure}
In this section, we describe our hardware and infrastructure that facilitated the large-scale pre-training of OneRec and introduce several optimizations that enhance training efficiency.

\noindent\textbf{Compute.} We utilize 90 servers for training, each equipped with 8 flagship GPUs and 2 CPUs interconnected via 400Gbps NVLink to ensure high-speed intra-node bandwidth.

\noindent\textbf{Networking.} Intra-node communication is managed by the efficient NVLink network, while inter-node communication is supported by 400Gbps RDMA for training traffic and 100Gbps TCP for training data and embedding prefetching operations.

\noindent\textbf{Storage.} Each server is equipped with 4 NVMe SSDs to expedite checkpoint writes, allowing for the storage of large-scale embedding parameters and dense parameters in HDFS with minimal downtime for fault tolerance.

\noindent\textbf{Training Acceleration.} For training acceleration, several core optimizations are implemented:

    \textit{1) Embedding Acceleration:}
    To manage the extensive embedding workload beyond CPU capacity, we utilize Kuaishou’s SKAI framework for GPU-based parameter servers. This framework leverages cross-GPU unified embedding tables, GPU caching paradigms, and prefetching pipelines to enhance training efficiency and reduce management overhead.

    \textit{2) Training Parallelism:}
    A combination of data parallelism, ZERO1~\citep{rajbhandari2020zero}, and gradient accumulation is employed for model training. ZERO1 is selected because the current model's dense parameters can be loaded on a single GPU, minimizing synchronization overhead in data parallel groups when interleaving multiple macro batches.
    
    \textit{3) Mixed Precision Training:}
    BFloat16 is used for computations in certain MLP networks to optimize performance.
    
    \textit{4) Compilation Optimization:}
    For attention networks, compilation optimizations are applied to reduce computational overhead.

Thanks to advancements in highly optimized training infrastructure, the model's training MFU has improved to 23.7\%, significantly narrowing the gap with the LLM training efficiency.

\subsection{Pre-training}

\paragraph{Pre-Training Data} As illustrated in Section~\ref{multiscale_feature}, our model takes multi-scale user behavior representations as input. The pre-training objective involves predicting sequences of target items for users. Each training sample comprises a target item which is tokenized into 3 semantic identifiers. This tokenization scheme results in 3 target tokens per training sample for the generative model's next-token prediction task. Our training pipeline processes approximately 18 billion samples daily, yielding a throughput of 54 billion tokens per day. The OneRec-0.935B model (detailed in Table~\ref{tab:arch-onerec}) achieves convergence after training on approximately 100 billion samples, corresponding to a total exposure of 300 billion tokens during pre-training.

\paragraph{Key Hyperparameters} 
The OneRec series comprises four models (two dense and two MoE variants) designed for recommendation tasks. Key architectural hyperparameters such as layer counts, hidden dimensions, and attention head numbers are detailed in Table~\ref{tab:arch-onerec}. In these models, encoders and decoders have the same number of layers. For dense variants, the standard Feed-Forward Networks (FFNs) typically expand the hidden dimension $d_{\text{ff}}$ to $2 \times d_{\text{model}}$. For the MoE variants, we replace standard FFNs with MoE layers in designated blocks, and employ SwiGLU FFNs ~\citep{thoppilan2022lamda, shazeer2020glu} as experts. Consistent with open-source MoE LLM settings ~\citep{jiang2024mixtral, fedus2022review}, the hidden dimension for each SwiGLU expert is calculated as $\frac{2}{3} \times 4 \times d_{\text{model}}$, ensuring it is a multiple of 128. 

The convergence curves for each model can be found in Section \ref{training_scaling}.

\begin{table}[H]
\caption{OneRec model architectures. "Layers" = \#Encoder + \#Decoder. "FFN Hid. Dim" is FFNs' intermediate size or MoEs' intermediate expert size.}
\small
\centering
\begin{tabular*}{\textwidth}{@{\extracolsep{\fill}}lcccccc@{}}
\toprule
Model & Layers & Hid. Dim & FFN Hid. Dim & Attn. Heads & Experts (Tot/Act) & MoE Loc. \\
\midrule
OneRec-0.015B (Dense) & 4 & 128 & 256 & 4 & N/A & N/A \\
OneRec-0.121B (Dense) & 8 & 1024 & 2048 & 8 & N/A & N/A \\
OneRec-0.935B (MoE) & 8 & 1024 & 2048 & 8 & 24 / 2 & Decoder \\
OneRec-2.633B (MoE) & 24 & 1024 & 2048 & 8 & 24 / 4 & Enc \& Dec \\
\bottomrule
\end{tabular*}
\label{tab:arch-onerec}
\end{table}

\subsection{Post-training}
In the post-training phase, we perform online training using real-time data streams. We simultaneously perform Reject Sampling Fine-Tuning (RSFT) and Reinforcement Learning (RL). For RSFT, we filter out the bottom 50\% of exposure sessions based on play duration. The training loss is the same as the $\mathcal{L}_{\rm NTP}$ loss in the pre-training process, but we apply annealing by reducing the learning rate of sparse parameters to $1 \times 10^{-4}$ and dense parameters to $8 \times 10^{-5}$. For RL, we randomly select 1\% of users from the RSFT data to generate RL samples.

\begin{figure}[t]
\centering
\includegraphics[width=0.95\textwidth]{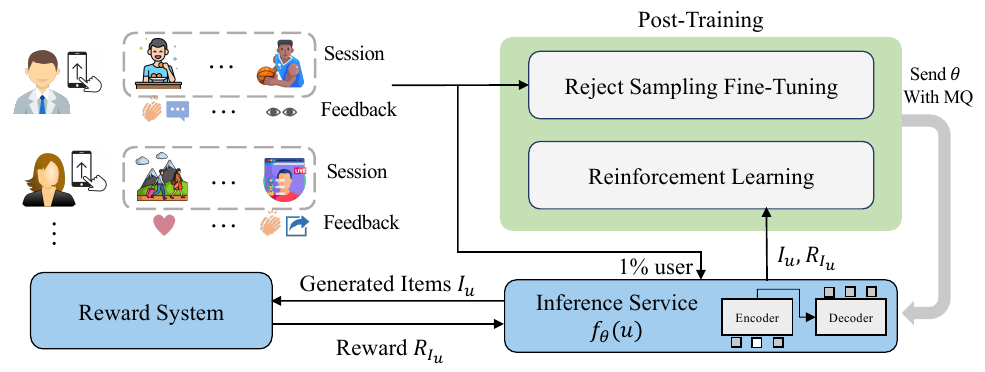}
\caption{The overall process of OneRec's post-training, including continual pre-training and reinforcement learning.}
\label{fig:post-training}
\end{figure}

To maximize computational resource utilization, we decouple the generation of RL samples from the training process by using an external inference service. During training, 1\% of users access the external service to generate 512 items, request rewards for each item from the reward model, and then return the data to the training task. The training task sends updated parameters to the external inference service via a Message Queue (MQ) every 1000 steps. The overall post-training process is summarized in Figure \ref{fig:post-training}.

\section{Evaluation}
\label{Evalution}

\subsection{Evaluation Metric}
We assess model performance through the following metrics:
\begin{itemize}[noitemsep,topsep=0pt,leftmargin=*]
    \item \textbf{Cross-entropy loss}: Next-token prediction loss $\mathcal{L}_{\mathrm{NTP}}$ curves.
    \item \textbf{P (preference)-Score}: Learned comprehensive evaluation metric, as detailed in Section~\ref{fusion_score}.
    \item \textbf{xtr metrics}: A set of user engagement indicators derived from a pre-trained ranking model~\citep{wang2024home, chang2023twin} currently deployed in our system, including:
    \begin{itemize}[noitemsep,topsep=0pt]
        \item \textbf{lvtr} (Long View Through Rate): Predicted probability of significant video viewing
        \item \textbf{vtr} (View Through Rate): Predicted probability of video viewing
        \item \textbf{ltr} (Like Through Rate): Predicted probability of video liking
        \item \textbf{wtr} (Follow Through Rate): Predicted probability of the creator following
        \item \textbf{cmtr} (Comment Through Rate): Predicted probability of video commenting
    \end{itemize}
\end{itemize}

For P-Score and xtr reward metrics, our evaluation system operates on streaming data where values may vary across different periods. Consequently, identical metrics may show different absolute values across experiments due to temporal variations in the data stream. However, we ensure reliable evaluation by conducting comparative experiments within the same periods and averaging results over sufficiently long observation windows, making our findings statistically confident.

\subsection{Scaling}
\subsubsection{Training Scaling} \label{training_scaling}

\paragraph{Parameters Scaling}

\begin{figure}[H]
    \centering
    \includegraphics[width=0.7\textwidth]{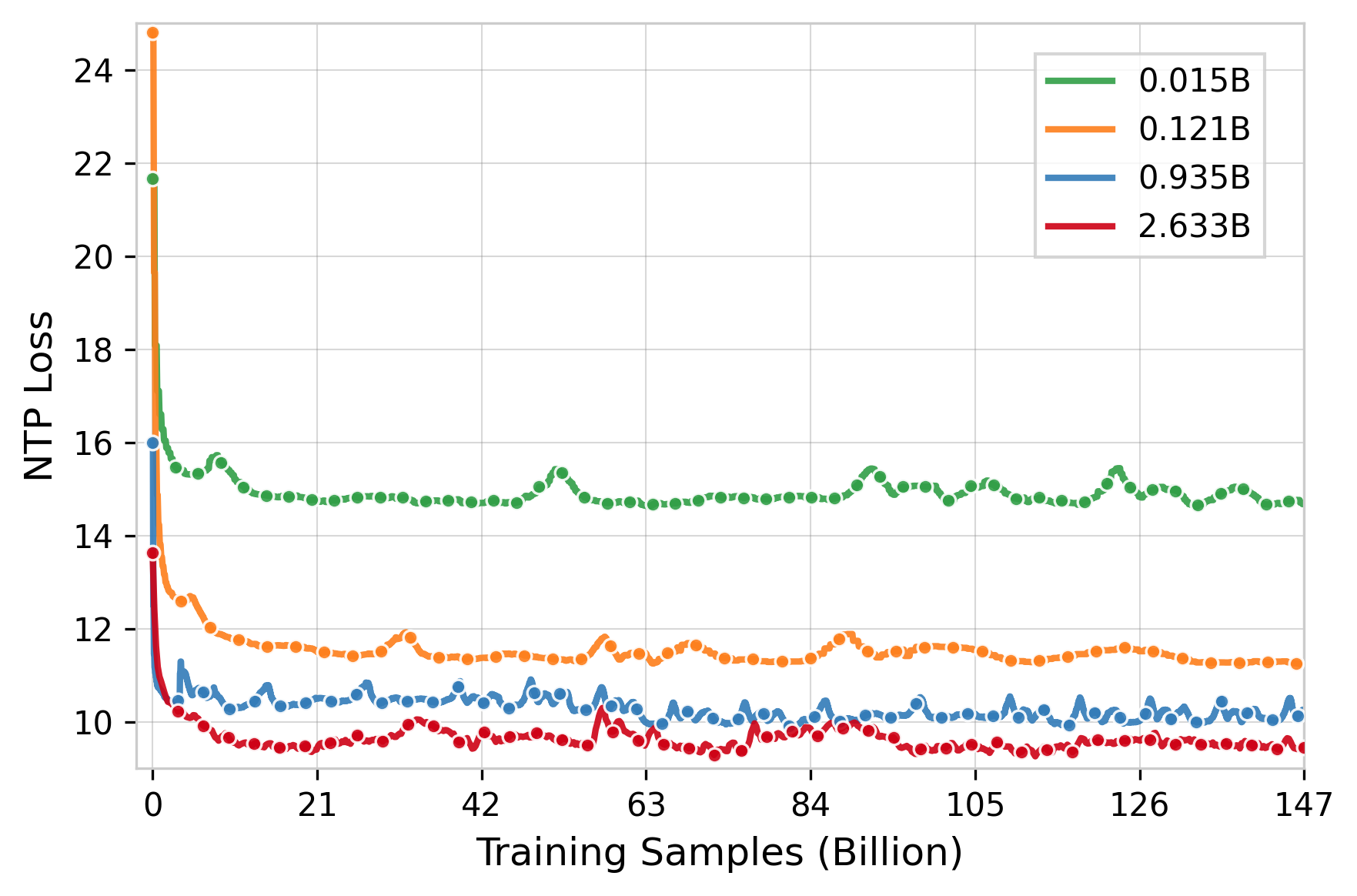}
    \caption{Comparison of loss curves for different OneRec model sizes, showing loss scaling with training samples.}
    \label{fig:loss_curve_comparison}
\end{figure}

The OneRec series includes models of varying sizes: OneRec-0.015B, OneRec-0.121B, OneRec-0.935B, and OneRec-2.633B, as detailed in Table~\ref{tab:arch-onerec}. We investigated the impact of model parameter count on performance. Figure~\ref{fig:loss_curve_comparison} illustrates the loss curves for these models, demonstrating a clear scaling trend where larger models achieve lower loss as training progresses. This indicates a strong capability for performance improvement with increased model size.

Regarding the influence of training data size, our experiments show that performance converges rapidly within the initial approximately 10 billion samples. While the rate of improvement diminishes significantly beyond this point, performance does not completely plateau and continues to benefit, albeit more slowly, from additional data (i.e., beyond 100 billion samples). This suggests that while substantial gains are achieved early in training, further, more gradual improvements are possible with larger datasets.

As model parameters scale up, load balancing among experts becomes a critical issue. Uneven expert utilization can lead to training inefficiency and suboptimal performance. We adopt DeepSeek's loss-free load balancing strategy \citep{liu2024deepseek}, which maintains expert utilization balance without introducing additional loss terms. With this strategy, we observe a loss reduction of 0.2, demonstrating its effectiveness in improving convergence for scaled OneRec models.

Beyond parameter scaling, we conduct additional experiments to validate the effectiveness of scaling across other key dimensions using our 0.935B model. These experiments encompass feature scaling (examining the impact of comprehensive feature engineering), codebook scaling (investigating the effect of vocabulary size expansion), and inference scaling (analyzing the influence of beam search parameters). Each dimension demonstrates distinct scaling behaviors and provides valuable insights for future model optimization.

\begin{figure}[H]
    \centering
    
    % --- 步骤 1: 将右侧的表格内容存入一个盒子，方便后续使用和测量 ---
    \newsavebox{\tablebox}
    \sbox{\tablebox}{
        \centering
        \footnotesize
        \begin{tabular}{lccr}
            \toprule
            \textbf{Metric} & \textbf{w/o. feature} & \textbf{w/. feature} & \textbf{Impr.} \\
            \midrule
            lvtr          & 0.4940 & 0.5500 & 11.34\% \\
            vtr           & 0.8730 & 0.8901 & 1.96\% \\
            ltr           & 0.0391 & 0.0441 & 12.79\% \\
            wtr           & 0.0190 & 0.0224 & 17.89\% \\
            cmtr          & 0.0919 & 0.1010 & 9.90\% \\
            P-score  & 0.0749    & 0.0966    & 28.88\% \\
            \bottomrule
        \end{tabular}
    }

    % --- 步骤 2: 创建两个 minipage，并使用 [b] 选项使它们底部对齐 ---
    % 左侧 minipage
    \begin{minipage}[b]{0.48\textwidth}
        \centering
        % 直接放置图片，它的高度将作为基准
        \includegraphics[width=\textwidth]{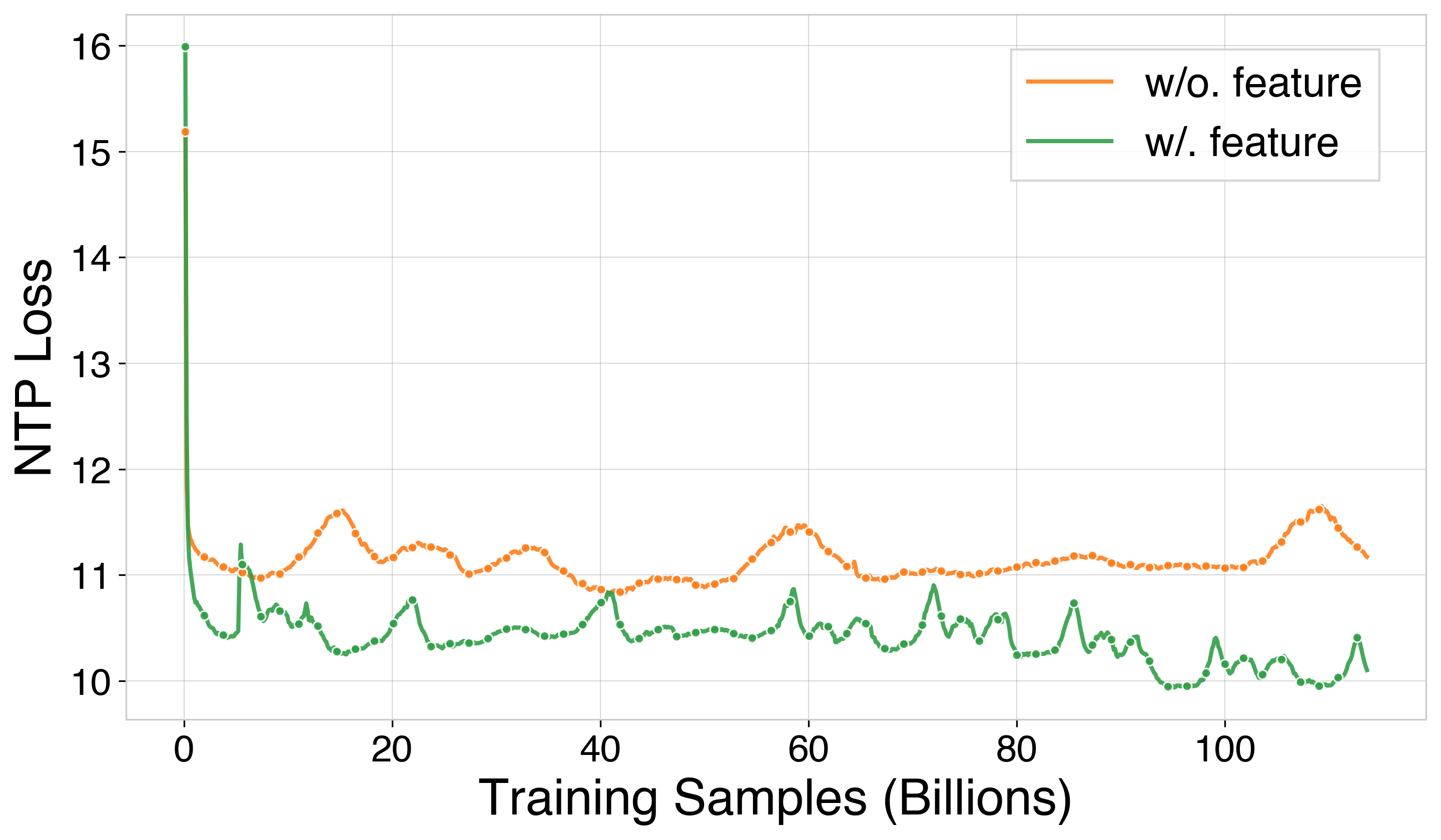}
        % 图片的 caption
        \captionof{figure}{Training loss comparison with and without additional features.}
        \label{fig:feature_scaling}
    \end{minipage}
    \hfill
    % 右侧 minipage
    \begin{minipage}[b]{0.48\textwidth}
        \centering
        % --- 步骤 3 (核心技巧): 创建一个与左侧图片等高的 parbox ---
        % \parbox[垂直对齐][高度][内部垂直对齐]{宽度}{内容}
        \parbox[c][\heightof{\includegraphics[width=\textwidth]{Figures/feature_scaling.png}}][c]{\linewidth}{
            \usebox{\tablebox} % 将之前保存的表格放入这个等高容器中
        }
        % 表格的 caption
        \captionof{table}{Performance comparison with and without additional features.}
        \label{tab:feature}
    \end{minipage}
\end{figure}

\paragraph{Feature Scaling} 
To investigate the impact of feature engineering on model performance, we compare the model with two input configurations: a baseline using only item ID \texttt{vid} embeddings from 256 positive-feedback items, and an enhanced version incorporating the comprehensive feature set described in our methodology. As shown in Figure~\ref{fig:feature_scaling} and Table~\ref{tab:feature}, the enhanced model with additional features achieves lower training loss and substantial improvements across multiple dimensions of recommendation quality.

\paragraph{Codebook Scaling} To investigate the impact of codebook size on model performance, we experiment by expanding the codebook from 8,192 to 32,768. It is important to note that NTP loss, as defined in our parameter scaling experiments, cannot be directly used for comparison here. This is because an increase in codebook size inherently expands the candidate set for the cross-entropy loss calculation, rendering direct loss comparisons misleading. Consequently, we evaluate performance using reward-based metrics. The performance improvements across various metrics are presented in Table~\ref{tab:codebook_scaling}. As shown in the result, increasing the codebook size yields significant improvements in playtime metrics and a slight gain in interaction metrics.

\paragraph{Infer Scaling}
We investigate the impact of different numbers of generated items in inference (Pass@K) on model performance. As detailed in Table~\ref{tab:infer_scaling_reformatted}, increasing K of Pass@K from 8 to 512 results in consistent performance improvements across all evaluated metrics. However, further increasing K from 512 to 1,024 yields only marginal gains. Considering the trade-off between performance improvements and the associated computational resource consumption, we select K=512 for deployment in our production environment.

\begin{figure}[H]
    \centering
    \begin{minipage}[t]{0.35\textwidth} % 缩小左边表格宽度
        \centering
        \scriptsize % 使用更小的字体
        \begin{tabular}{@{}lccr@{}} % @{} removes padding at table edges
            \toprule
            \textbf{Metric} & \textbf{Size=8K} & \textbf{Size=32K} & \textbf{Impr.} \\
            \midrule
            lvtr         & 0.5118 & 0.5245 & 2.48\% \\
            vtr          & 0.9384 & 0.9491 & 1.14\% \\
            ltr          & 0.0298 & 0.0299 & 0.34\% \\
            wtr          & 0.0153 & 0.0154 & 0.65\% \\
            cmtr         & 0.0650 & 0.0664 & 2.15\% \\
            P-score  & 0.2516    & 0.2635    & 4.75\% \\
            \bottomrule
        \end{tabular}
        \captionof{table}{Codebook Scaling.}
        \label{tab:codebook_scaling}
    \end{minipage}
    \hfill
    \begin{minipage}[t]{0.6\textwidth} % 调整右边表格宽度
        \centering
        \scriptsize % 使用更小的字体以节省空间
        \begin{tabular}{@{}lccccc@{}} % 去掉最后的r，减少右边距
            \toprule
            \textbf{Metric} & \textbf{Pass@8} & \textbf{Pass@64} & \textbf{Pass@512} & \textbf{Pass@1024} & \textbf{Impr.} \\
            \midrule
            lvtr         & 0.3675 & 0.4927 & 0.5351 & 0.5443 & 48.11\% \\
            vtr          & 0.9444 & 0.9462 & 0.9513 & 0.9530 & 0.91\% \\
            ltr          & 0.0278 & 0.0346 & 0.0425 & 0.0452 & 62.59\% \\
            wtr          & 0.0114 & 0.0138 & 0.0182 & 0.0197 & 72.81\% \\
            cmtr         & 0.0350 & 0.0566 & 0.0809 & 0.0891 & 154.57\% \\
            P-score  & 0.0811    & 0.2051    & 0.3375   & 0.3859   & 376.10\% \\
            \bottomrule
        \end{tabular}
        \captionof{table}{Inference Pass@K Scaling.}
        \label{tab:infer_scaling_reformatted}
    \end{minipage}
\end{figure}
\subsubsection{Semantic Identifier Input Representation}
As model sizes scale to billions of parameters, we explore an alternative input representation strategy that leverages video semantic identifiers for user interaction histories instead of constructing separate sparse embeddings for video identifiers (\texttt{vid}). This semantic identifier input achieves performance comparable to traditional sparse embedding methods, while offering significant advantages in parameter efficiency, communication overhead, and sequence processing capacity that make it particularly promising for further scaling exploration.

\begin{figure}[H]
    \centering
    
    % --- 步骤 1: 将新的表格内容存入盒子 ---
    % \newsavebox{\tablebox}
    \sbox{\tablebox}{
        \centering
        \footnotesize
        \begin{tabular}{lccc}
            \toprule
            \textbf{Metric} & \textbf{VID} & \textbf{Semantic ID} & \textbf{Impr.} \\
            \midrule
            lvtr & 0.4447 & 0.4467 & 0.45\% \\
            vtr & 0.8725 & 0.8726 & 0.01\% \\
            ltr & 0.0336 & 0.0336 & 0.00\% \\
            wtr & 0.0104 & 0.0105 & 0.96\% \\
            cmtr & 0.0565 & 0.0573 & 1.42\% \\
            P-score  & 0.0371 & 0.0378 & 1.74\% \\
            \bottomrule
            \end{tabular}
    }

    % --- 步骤 2: 创建两个底部对齐 [b] 的 minipage ---
    % 左侧 minipage
    \begin{minipage}[b]{0.48\textwidth}
        \centering
        % 放置新图片
        \includegraphics[width=\textwidth]{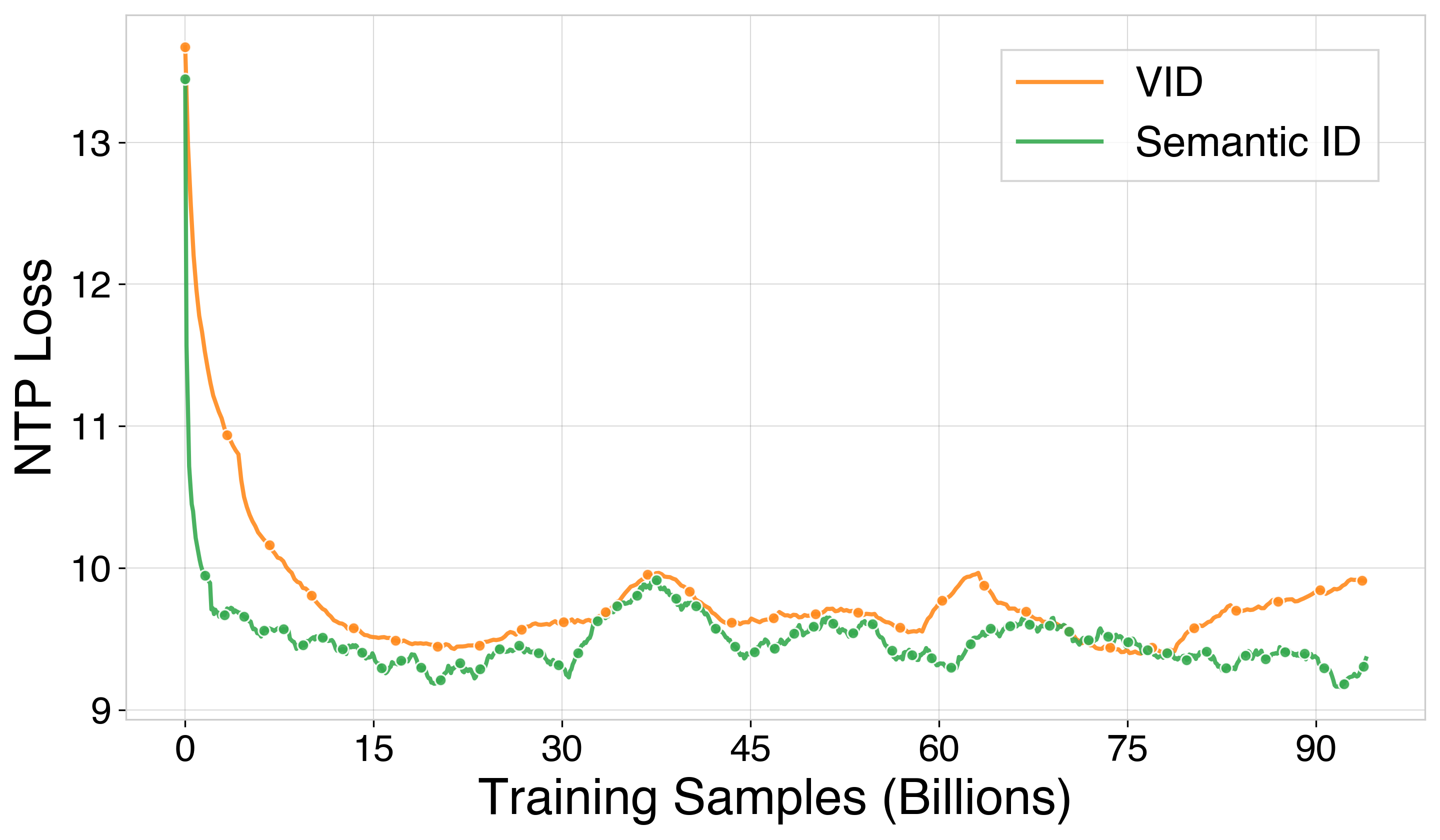}
        % 图片的 caption
        \captionof{figure}{Training loss comparison between OneRec-2.633B with semantic identifier input and sparse embedding input.}
        \label{fig:sid}
    \end{minipage}
    \hfill
    % 右侧 minipage
    \begin{minipage}[b]{0.48\textwidth}
        \centering
        % --- 步骤 3: 创建一个与新图片等高的 parbox，并垂直居中表格 ---
        \parbox[c][\heightof{\includegraphics[width=\textwidth]{Figures/sid_vs_pid.png}}][c]{\linewidth}{
            \usebox{\tablebox} % 使用新表格
        }
        % 表格的 caption
        \captionof{table}{Performance comparison between OneRec-2.633B with semantic identifier input and sparse embedding input.} 
        \label{tab:sid}
    \end{minipage}
\end{figure}

\paragraph{Scaling Performance Analysis}
\label{sec:sid}
As shown in Figure~\ref{fig:sid}, our empirical analysis reveals that at scale (2.6B parameters), the semantic identifier input approach achieves performance comparable to or exceeding traditional sparse embedding methods. 

\paragraph{Advantages and Future Scaling}

The semantic identifier approach provides several key advantages over traditional sparse embedding methods, making it particularly attractive for further scaling exploration:
\begin{itemize}
    \item \textbf{Parameter Efficiency:} By sharing embeddings between input and output representations, the model eliminates the need for separate sparse embedding tables for \texttt{vid}. This dramatically reduces the total parameter count, particularly for Kuaishou with billions of items.
    \item \textbf{Communication Efficiency:} In distributed training environments, sparse embedding operations require extensive parameter server communication for embedding lookup and gradient updates. The semantic identifier approach reduces communication overhead by leveraging dense operations and shared vocabulary, leading to faster training throughput and reduced communication bottlenecks.
    \item \textbf{Extended Sequence Capacity:} The elimination of large sparse embedding tables enables the allocation of computational resources toward processing longer user interaction sequences. This allows the model to capture more comprehensive user preference evolution patterns, potentially extending sequence lengths from thousands to tens of thousands of interactions.
    \item \textbf{Representation Consistency:} Sharing the same semantic space between input and output ensures representational consistency and enables the model to learn more coherent item-to-item relationships. This unified representation has the potential to facilitate better generalization across different recommendation scenarios.
\end{itemize}

Given these compelling advantages and the competitive performance demonstrated at the 2.6B parameter scale, we are actively pursuing further scaling exploration based on semantic identifier input representation. This approach promises to unlock new possibilities for large-scale recommendation systems while maintaining computational efficiency and architectural elegance.
\subsection{Reinforcement Learning}

\subsubsection{User Preference Alignment}
Defining what constitutes a "good" recommendation has always been a challenging task.
% Our reward system employs the multi-objective P-Score (Preference Score) to evaluate the overall quality of a recommended video, which inevitably causes a seesaw effect among certain metrics, obscuring the true effectiveness of RL. 
To rigorously verify RL's impact, we use the single-objective \textit{vtr} (view-through rate) as the reward, which corresponds to online metrics such as Watch Time and App Stay Time. The reported online results are relative improvements compared to Kuaishou's traditional recommendation system, referred to as the overall baseline. \textit{Relative Impr.} in the table indicates the relative enhancement of the latter group over the former group.

Notably, while using \textit{vtr} as the reward can significantly improve duration metrics, it does not necessarily indicate a high-quality recommendation, as other metrics, such as Video View, which represent the number of videos viewed, may decrease significantly. We primarily focus on Watch Time and App Stay Time to find the optimal RL setting, and ultimately use it to validate the benefits of the P-Score reward.

\paragraph{Sampling Efficiency}
Reinforcement learning optimizes the probability distribution of sampled items to increase the likelihood of selecting high-reward items, thereby significantly enhancing sampling efficiency. To quantify this effect, we conduct multi-point sampling experiments at pass@32, pass@128, and pass@512, with results summarized in Table \ref{tab:infer_bs}. Treating the model without RL as the baseline, we define the improvement in app stay time as the sampling efficiency gap. Notably, RL shows the most substantial improvement gap at pass@32, indicating that the accuracy of top-ranked items is significantly enhanced. This improvement is crucial for reducing sampling overhead, as it ensures high precision when sampling a small number of items. In recommendation systems, balancing cost and benefit is essential, and the enhanced accuracy at lower sample numbers \( K \) provides a solid foundation for achieving this balance.

\begin{table}[h]
    \centering
    \begin{tabular}{llllll}
    \toprule
    & Method & vtr & Watch time & App Stay Time & \textcolor{gray}{Video View$^1$} \\
    \midrule
    \multirow{3}{*}{Pass@32} 
    & OneRec w/o RL & 0.1978 & +1.62\% & -0.10\% & \textcolor{gray}{-4.18\%} \\
    & OneRec w/ RL & 0.2138 & +3.17\% & +0.39\% & \textcolor{gray}{-9.87\%} \\
    & Relative Impr. & +8.08\% & +1.55\% & +0.49\%$\uparrow\uparrow\uparrow$ & \textcolor{gray}{-3.69\%} \\
    \midrule
    \multirow{3}{*}{Pass@128} 
    & OneRec w/o RL & 0.2239 & +4.61\% & +1.11\% & \textcolor{gray}{-12.75\%} \\
    & OneRec w/ RL & 0.2387 & +5.22\% & +1.49\% & \textcolor{gray}{-15.06\%} \\
    & Relative Impr. & +6.61\% & +1.53\% & +0.38\%$\uparrow\uparrow$ & \textcolor{gray}{-2.65\%} \\
    \midrule
    \multirow{3}{*}{Pass@512} 
    & OneRec w/o RL & 0.2444 & +6.32\% & +1.66\% & \textcolor{gray}{-15.54\%} \\
    & OneRec w/ RL & 0.2494 & +5.88\% & +1.75\% & \textcolor{gray}{-13.88\%} \\
    & Relative Impr. & +2.05\% & -0.41\% & +0.09\%$\uparrow$ & \textcolor{gray}{+1.97\%} \\
    \bottomrule
    \end{tabular}
    \caption{The impact of reinforcement learning under different numbers of generated items (Pass@K) during inference.}
    \label{tab:infer_bs}
\end{table}

\footnotetext[1]{Video View is provided for reference only, as our primary focus is on Watch Time and App Stay Time to determine the optimal RL setting.}

\paragraph{Search Space}
In ECPO training, expanding the action search space increases the likelihood of discovering the optimal item with maximum reward, albeit at higher computational costs. To investigate this trade-off, we examine how the search space size (i.e., group size) affects performance. The results for pass@128 are summarized in Table \ref{tab:search_space}. From Table \ref{tab:search_space}, we observe a significant improvement in performance when the group size is increased from 128 to 512. This clearly demonstrates the positive impact of expanding the search space. It is somewhat disappointing that increasing the search space to 2048 does not yield much additional benefit, which might be due to the current reference model's diversity not being sufficient to discover more and better items. Nonetheless, this finding is promising, and we empirically suggest setting the ECPO training group size to approximately four times the inference output quantity for optimal results.

\begin{table}[h]
    \centering
    \begin{tabular}{lccc>{\columncolor{white}\color{gray}}c}
    \toprule
    Group Size & vtr & Watch time & App Stay Time & Video View$^1$ \\
    \midrule
    0(w/o RL) & 0.2198 & +4.61\% & +1.11\% & \textcolor{gray}{-12.75\%} \\
    128 & 0.2303 & +5.22\% & +1.49\% & \textcolor{gray}{-15.06\%} \\
    512 & 0.2350 & +5.73\% & +1.82\% & \textcolor{gray}{-15.49\%} \\
    2048 & 0.2352 & +5.84\% & +1.78\% & \textcolor{gray}{-15.49\%} \\
    \bottomrule
    \end{tabular}
    \caption{Performance of different group sizes when calculating ECPO loss on Pass@128.}
    \label{tab:search_space}
\end{table}
\paragraph{Search Strategy}

Reinforcement learning for large language models typically employs top-$k$ and top-$p$ sampling for sample generation. In OneRec, we also explore beam search as an alternative strategy. Table~\ref{tab:sampling_strategy} compares the results of these two approaches, revealing that beam search significantly outperforms top-$k$ and top-$p$ sampling in OneRec's reinforcement learning framework. This improvement stems from the inherent regularity of semantic ID structures, which follow a prefix tree encoding scheme and thus align well with the systematic exploration of beam search.
\begin{table}[h]
    \centering
    \begin{tabular}{lccc>{\columncolor{white}\color{gray}}c}
    \toprule
    & vtr & Watch time & App Stay Time & Video View$^1$ \\
    \midrule
    Top-$k$+Top-$p$ & 0.2131 & +4.45\% & +1.16\% & \textcolor{gray}{-13.61\%} \\
    Beam Search & 0.2162 & +5.35\% & +1.76\% & \textcolor{gray}{-13.30\%} \\
    Relative Impr. & +1.45\% & +0.87\% & +0.60\% & \textcolor{gray}{+0.36\%} \\
    \bottomrule
    \end{tabular}
    \caption{Performance of reinforcement learning with different search strategies.}
    \label{tab:sampling_strategy}
\end{table}

\paragraph{Reference Model}
In this section, we compare two reference models for strategy generation in ECPO: (1) the pre-trained model (off-policy) and (2) the current policy model (on-policy). The experimental results are summarized in Table \ref{tab:ref_model}. From the table, it is evident that using the current policy model yields better results, especially in offline reward evaluation. This indicates that the on-policy approach allows the model to continuously teach itself, breaking through the limitations of the reference model and achieving a higher upper limit. However, in terms of online performance, the improvement with the on-policy approach is not very significant. This is due to the suboptimal definition of the reward, leading to slight reward hacking. We will focus on this aspect as a key direction for future work.
\begin{table}[h]
    \centering
    \begin{tabular}{lccc>{\columncolor{white}\color{gray}}c}
    \toprule
    Reference Model & vtr & Watch time & App Stay Time & Video View$^1$ \\
    \midrule
    Pre-trained Model & 0.2262 & +5.35\% & +1.51\% & \textcolor{gray}{-13.51\%} \\
    Current Policy Model & 0.2389 & +6.19\% & +1.56\% & \textcolor{gray}{-13.89\%} \\
    Relative Impr. & +5.61\% & +0.79\% & +0.04\% & \textcolor{gray}{-13.89\%} \\
    \bottomrule
    \end{tabular}
    \caption{Performance of reinforcement learning with different reference models.}
    \label{tab:ref_model}
\end{table}

\paragraph{P-Score Reward}
In this section, we observe the comprehensive improvements achieved through reinforcement learning when using P-Score as the reward. Based on the conclusions from the above ablation experiments, we select the optimal RL setting, which involves using beam search for RL sample generation and employing the current policy model as the reference model. We examine the impact of RL in two scenarios, including Kuaishou and Kuaishou Lite, with the results summarized in Table 1. From the table, we can conclude that in both scenarios, P-Score significantly improves App Stay Time and Watch Time while also increasing Video View, indicating an enhancement in the overall user recommendation experience. 

\begin{table}[h]
    \centering
    \begin{tabular}{lccc}
    \toprule
    Scenario  & Watch time & App Stay Time & Video View \\
    \midrule
    Kuaishou  & +0.21\% & +0.26\% & +0.17\% \\
    Kuaishou Lite & +0.71\%  & +0.22\% & +0.35\% \\
    \bottomrule
    \end{tabular}
    \caption{The relative improvement of OneRec with P-Score Reward compared to without it in the Kuaishou and Kuaishou Lite scenarios.}
    \label{tab:p_reward}
\end{table}
\subsubsection{Generation Format Regularization}
In this section, we conduct experiments to verify the effectiveness of format reward. As mentioned in Section \ref{sec:format}, after incorporating reinforcement learning into the pre-trained model, the legality of the model's output significantly drops to below 50\% due to the squeezing effect. This means that more than half of the generated semantic IDs do not correspond to actual video IDs, which is detrimental to the stability of recommendations and the scalability of inference.
We evaluate the impact of format reward by comparing two sample selection methods for computing format loss: (1) selecting the top-5 highest-probability samples from 128 generated candidates, and (2) randomly selecting 5 samples. 

\begin{figure}[ht]
\centering
\includegraphics[width=.42\textwidth]{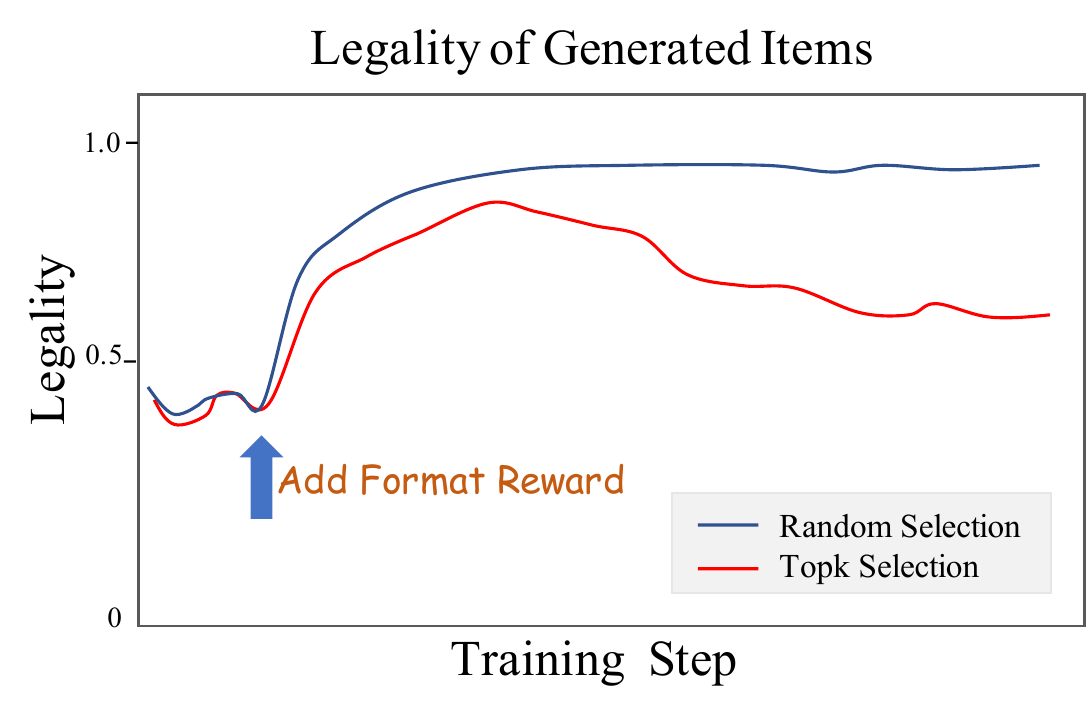}
\includegraphics[width=.42\textwidth]{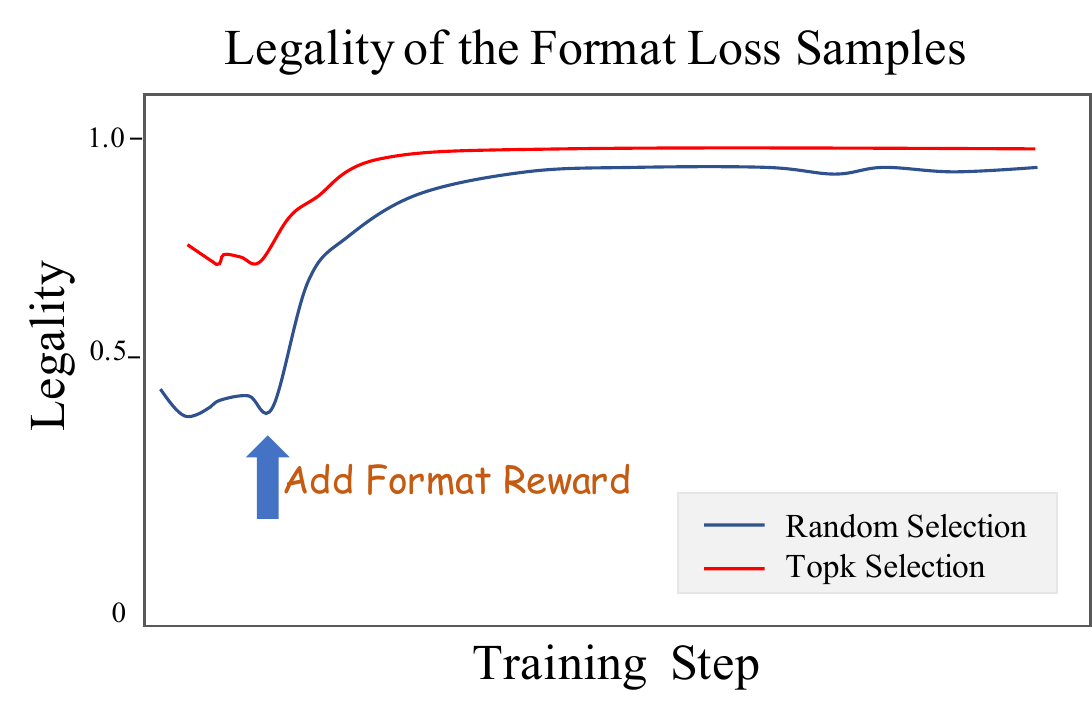}
\caption{The impact of training with format reward with samples obtained through different sampling strategies on the model's legality.}
\label{fig:format}
\end{figure}

Figure~\ref{fig:format} illustrates their effects on output legality. The left figure shows legality rates across all 128 generated samples, while the right panel focuses on the selected samples. Without format rewards, baseline legality remains below 50\%. The Top-k Selection approach produces an interesting pattern: while overall legality initially rises then falls, the selected samples rapidly achieve 100\% legality, suggesting the model learns to generate legal outputs only within the top-ranked subset. In contrast, Random Selection presents a more challenging learning objective, yet drives steady improvement - ultimately reaching 95\% legality without showing a decline.

Notably, format reward integration yields benefits beyond legality alone. Online metrics demonstrate substantial gains: +0.13\% in APP Stay Time and +0.30\% in Watch Time. This experimental case not only validates the format reward mechanism but also highlights the critical role of careful reward design in reinforcement learning systems.

\subsubsection{Industrial Scenario Alignment}
\label{sec:sir_exp}
In this section, we present a practical example of using reinforcement learning to address industrial challenges. On the Kuaishou platform, viral content farms represent a significant portion of content creators, primarily producing repurposed and clipped videos with inconsistent quality. While OneRec demonstrates superior performance over traditional recommendation systems across multiple business metrics, we observe that without proper post-filtering strategies, the exposure ratio of viral content increases significantly, which may negatively impact the platform's ecosystem. 

The optimal proportion of viral content videos can be set to $f$. When the proportion exceeds $f$, we down-weight their P-score reward to suppress them while maintaining the system's perception of the quality of these contents.
\begin{equation}
    r_i' = \begin{cases} 
    r_i & \text{if } o_i\notin I_\text{viral} \\
    \alpha r_i & \text{if } o_i\in I_\text{viral}
    \end{cases},
\end{equation}
where $\alpha \in (0,1)$ is the suppression factor.

We term this approach Specific Industrial Reward (SIR). Experimental results show that SIR effectively reduces viral content exposure by 9.59\% while maintaining stable performance on core metrics (Watch time and APP Stay Time). This experiment highlights OneRec's key advantage: the ability to achieve precise and consistent optimization through reinforcement learning's reward-shaping capability, a feature fundamentally unavailable in traditional recommendation systems.

\subsection{Tokenizer}
\label{exp:tokenizer}

We employ three metrics to comprehensively evaluate our tokenization method, encompassing aspects of accuracy, resource utilization, and distribution uniformity:

\begin{itemize}[noitemsep,topsep=0pt,leftmargin=*]
\item \textbf{Reconstruction Loss}: This metric assesses the accuracy with which discrete tokens reconstruct the original input, serving as an indicator of the model's fidelity in preserving the input data.
\item \textbf{Codebook Utilization \citep{zhu2024scaling}}: This metric evaluates the efficiency of vector usage within the codebook, reflecting how effectively the model leverages available resources to represent data.
\item \textbf{Token Distribution Entropy \citep{bentz2016word}}: Utilizing Shannon entropy, this metric quantifies the uniformity of token distribution, providing insight into the diversity and balance of token allocation across the model.
\end{itemize}

\begin{table}[h]
\caption{Performance comparison of tokenization algorithms with a three-layer 8,192 codebook.}
\label{tab:tokenization}
\centering
\begin{tabular}{l|c|c|c}
\hline
\multicolumn{2}{l|}{} & RQ-VAE & RQ-Kmeans \\
\hline
\multicolumn{2}{l|}{Reconstruction Loss $\downarrow$} & 0.0548 & \textbf{0.0410} \\
\hline
\multirow{3}{*}{Codebook Utilization $\uparrow$} & layer 1 & \textbf{1.0000} & \textbf{1.0000} \\
 & layer 2 & 0.9963 & \textbf{1.0000} \\
 & layer 3 & 0.9958 & \textbf{1.0000} \\
\hline
\multirow{3}{*}{Token Distribution Entropy $\uparrow$} & layer 1 & 8.3892 & \textbf{8.9191} \\
 & layer 2 & 8.4805 & \textbf{8.7770} \\
 & layer 3 & 8.6037 & \textbf{8.7276} \\
\hline
\end{tabular}
\end{table}

As shown in Table~\ref{tab:tokenization}, compared to RQ-VAE, RQ-Kmeans's reconstruction loss is reduced by 25.18\%, demonstrating superior accuracy in preserving input information. Simultaneously, RQ-Kmeans achieves perfect utilization (1.0000) in all three layers, indicating optimal resource efficiency in the codebook, while RQ-VAE shows slightly lower utilization rates in layers 2 and 3. Furthermore, RQ-Kmeans exhibits higher entropy values in all three layers compared to RQ-VAE, with significant improvements of 6.31\%, 3.50\%, and 1.44\% in layers 1, 2, and 3, respectively, suggesting that RQ-Kmeans produces a more uniform and balanced token distribution, which is beneficial for model stability and generalization capability. These comprehensive results demonstrate that RQ-Kmeans outperforms RQ-VAE across all three evaluation metrics, making it a more effective choice for tokenization.

Further qualitative analyses of item representation and tokenization quality are provided in Appendix~\ref{appendix:tokenizator}.

\subsection{Online A/B Test}
\label{online_ab_test}

We deployed OneRec in two major short-video scenarios on Kuaishou: the main Kuaishou feed and Kuaishou Lite feed - the platform's highest-traffic scenarios with daily active users of 400 million. Using a 5\% traffic experimental group observed over one week, our primary metrics were APP Stay Time (reflecting total user engagement time) and LT7 (7-day Lifetime). Two experimental groups were established: one employing a pure generative model (OneRec) and another augmenting generative outputs with reward model based selection (OneRec with RM Selection). As shown in Table \ref{online_ab}, the pure generative model with RL-based user preference alignment remarkably matched the performance of the entire complex recommendation system. Further applying reward model selection achieved statistically significant \textbf{improvements of +0.54\% and +1.24\% in APP Stay Time, and +0.05\% and +0.08\% in LT7} on these two scenarios, respectively. Notably, improvements of \textbf{0.1\% in APP Stay Time and 0.01\% in LT7} are already considered statistically significant on Kuaishou. Additionally, OneRec demonstrated significant gains across all interaction metrics (likes, follows, comments, etc.), indicating its ability to converge multi-task systems to a more balanced equilibrium without seesaw effects. After validation, we've expanded deployment to approximately 25\% of total QPS, with implementation details available in Appendix~\ref{appendix:online_ab}.

In addition to Kuaishou's short video recommendation scenarios, experiments have also been conducted in one of its significant business scenes — Local Life Service. The results demonstrate that OneRec achieves \textbf{a 21.01\% growth in GMV, a 17.89\% increase in order volume, an 18.58\% rise in buyer numbers, and a 23.02\% increase in new buyer acquisition}. Consequently, the system has now taken over \textbf{100\% of QPS} for this business scenario. After full deployment, we observe even stronger growth across all metrics compared to the initial experimental phase. These results prove OneRec's generalizability across diverse business contexts for enhanced recommendation performance.

\begin{table*}[h]
    \centering
    \caption{The absolute improvement of OneRec compared to the current multi-stage system in the online A/B testing setting.}
    \renewcommand{\arraystretch}{1.5} % 行高
    \setlength{\tabcolsep}{8pt}      % 列间距
    \begin{tabular}{p{3cm}|p{3.3cm}|p{3.3cm}|p{4.4cm}}
    \toprule
    Scenarios & Online Metrics &  OneRec &  OneRec with RM Selection\\
    \hline
    %\midrule
    \multirow{8}{*}{Kuaishou}&App Stay Time &+0.01\% &+0.54\%\\
    %&LT7 & &+0.05\%\\
    &Watch Time &+0.07\%&+1.98\%\\
    &Video View &+1.98\%&+2.52\% \\
    % &Engagement Depth& & \\
    &Like &-2.00\%&+2.43\%\\
    &Follow&-2.88\%&+3.24\%\\
    &Comment &-1.56\%&+5.27\% \\
    &Collect &-0.61\%&+2.93\%\\
    &Foward &+0.27\%&+5.90\%\\
    \hline
    \multirow{8}{*}{Kuaishou Lite}&App Stay Time & +0.06\% & +1.24\%\\
    %&LT7 & &+0.08\%\\
    &Watch Time &+0.05\% & +3.28\%\\
    &Video View &+2.40\% & +3.39\%\\
    % &Engagement Depth& & \\
    &Like &-2.64\% &+1.49\%\\
    &Follow &-2.75\% & +2.28\%\\
    &Comment &-2.23\% & +3.20\% \\
    &Collect &-1.76\% & +1.91\%\\
    &Foward &-1.86\% &+3.48\% \\
    \bottomrule
    \end{tabular}
    \label{online_ab}
\end{table*}

\noindent\textbf{Infrastructure and Efficiency}
We utilize  NVIDIA L20 GPUs for inference, and each server is equipped with 4 GPUs and 2 CPUs, connected via PCIe. We adopt Kuaishou's prediction platform - UniPredict to support online traffic. The inference service and embedding service are deployed in a 200Gb RDMA data center, leveraging RoCE networking. The maximum inter-machine communication bandwidth reaches 800Gb.
In order to improve the efficiency, we employ TensorRT to compile and optimize the model's computation graph. Through custom plugins, we achieve high-performance implementations of cross-attention, MoE, and other operations. Combined with batching and MPS techniques, we achieve a 5 $\times$ throughput improvement, reaching an MFU of 28.8\%.

\section{Conclusion, Limitations, and Future Directions}
\label{Conclusion}

In this paper, we introduce OneRec, a novel end-to-end generative recommendation architecture. Built as an encoder-decoder model, it compresses users’ lifelong behavior sequences via its encoder to derive user interests, while leveraging Mixture-of-Experts (MoE) to massively scale decoder parameters for precise short-video recommendation decoding. During post-training, we develop a customized reinforcement learning (RL) framework to refine recommendations by aligning model outputs with the reward function. Thanks to meticulous engineering optimizations, OneRec achieves 23.7\% and 28.6\% Model FLOPs Utilization (MFU) in training and inference — a dramatic improvement from single-digit baselines — closing the gap with the mainstream AI community. Notably, this compute-intensive design operates at 10.6\% the OPEX of conventional recommender systems. Comprehensive evaluations demonstrate that OneRec has surpassed existing recommendation systems in both effectiveness and efficiency. While acknowledging its powerful performance and high cost-effectiveness, we also recognize some limitations of OneRec and plan to strategically invest in the following areas:
\begin{itemize}[noitemsep,topsep=0pt,leftmargin=*]
    \item Inference Stage Scaling: The step scaling during the inference phase is not yet apparent, indicating that OneRec currently lacks strong reasoning capabilities.
    \item Multimodal Integration: OneRec has not yet integrated with LLMs (Large Language Models) and VLMs (Vision Language Models). User behavior is also a modality, and in the future, we plan to design solutions that allow user behavior modality to become a native multimodal model, similar to vision and audio alignment.
    \item Reward System Design: The reward system design is still very rudimentary, which is an exciting aspect. Historically, recommendation systems were not end-to-end, making it difficult to define and iterate on what constitutes a good recommendation result. Under the OneRec architecture, the reward system impacts both online results and offline training. We believe that the structure will soon lead to technological breakthroughs in the reward system for recommendations.
\end{itemize}

OneRec establishes an entirely new architecture, introducing a transformative framework for technological evolution, business value optimization, and team collaboration. While currently not yet deployed across all traffic scenarios in Kuaishou, we have adopted this as our foundational approach to systematically push the boundaries of algorithmic innovation while refining team collaboration mechanisms, thereby building scalable infrastructure capable of supporting traffic growth at scale.
\newpage

% Bibliography components
\bibliographystyle{abbrvnat}
\nobibliography*
\bibliography{bibtex}

\begin{thebibliography}{36}
\providecommand{\natexlab}[1]{#1}
\providecommand{\url}[1]{\texttt{#1}}
\expandafter\ifx\csname urlstyle\endcsname\relax
  \providecommand{\doi}[1]{doi: #1}\else
  \providecommand{\doi}{doi: \begingroup \urlstyle{rm}\Url}\fi

\bibitem[Bentz and Alikaniotis(2016)]{bentz2016word}
C.~Bentz and D.~Alikaniotis.
\newblock The word entropy of natural languages.
\newblock \emph{arXiv preprint arXiv:1606.06996}, 2016.

\bibitem[Cao et~al.(2025)Cao, Xu, Cheng, Guo, Tang, Wang, Leng, Yang, Liu, Niu, et~al.]{cao2025pantheon}
J.~Cao, P.~Xu, Y.~Cheng, K.~Guo, J.~Tang, S.~Wang, D.~Leng, S.~Yang, Z.~Liu, Y.~Niu, et~al.
\newblock Pantheon: Personalized multi-objective ensemble sort via iterative pareto policy optimization.
\newblock \emph{arXiv preprint arXiv:2505.13894}, 2025.

\bibitem[Chang et~al.(2023)Chang, Zhang, Fu, Zang, Guan, Lu, Hui, Leng, Niu, Song, et~al.]{chang2023twin}
J.~Chang, C.~Zhang, Z.~Fu, X.~Zang, L.~Guan, J.~Lu, Y.~Hui, D.~Leng, Y.~Niu, Y.~Song, et~al.
\newblock Twin: Two-stage interest network for lifelong user behavior modeling in ctr prediction at kuaishou.
\newblock In \emph{Proceedings of the 29th ACM SIGKDD Conference on Knowledge Discovery and Data Mining}, pages 3785--3794, 2023.

\bibitem[Cheng et~al.(2016)Cheng, Koc, Harmsen, Shaked, Chandra, Aradhye, Anderson, Corrado, Chai, Ispir, et~al.]{cheng2016wide}
H.-T. Cheng, L.~Koc, J.~Harmsen, T.~Shaked, T.~Chandra, H.~Aradhye, G.~Anderson, G.~Corrado, W.~Chai, M.~Ispir, et~al.
\newblock Wide \& deep learning for recommender systems.
\newblock In \emph{Proceedings of the 1st workshop on deep learning for recommender systems}, pages 7--10, 2016.

\bibitem[Dubey et~al.(2024)Dubey, Jauhri, Pandey, Kadian, Al-Dahle, Letman, Mathur, Schelten, Yang, Fan, et~al.]{dubey2024llama}
A.~Dubey, A.~Jauhri, A.~Pandey, A.~Kadian, A.~Al-Dahle, A.~Letman, A.~Mathur, A.~Schelten, A.~Yang, A.~Fan, et~al.
\newblock The llama 3 herd of models.
\newblock \emph{arXiv preprint arXiv:2407.21783}, 2024.

\bibitem[Fedus et~al.(2022)Fedus, Dean, and Zoph]{fedus2022review}
W.~Fedus, J.~Dean, and B.~Zoph.
\newblock A review of sparse expert models in deep learning.
\newblock \emph{arXiv preprint arXiv:2209.01667}, 2022.

\bibitem[Grattafiori et~al.(2024)Grattafiori, Dubey, Jauhri, Pandey, Kadian, Al-Dahle, Letman, Mathur, Schelten, Vaughan, et~al.]{grattafiori2024llama}
A.~Grattafiori, A.~Dubey, A.~Jauhri, A.~Pandey, A.~Kadian, A.~Al-Dahle, A.~Letman, A.~Mathur, A.~Schelten, A.~Vaughan, et~al.
\newblock The llama 3 herd of models.
\newblock \emph{arXiv preprint arXiv:2407.21783}, 2024.

\bibitem[Guo et~al.(2017)Guo, Tang, Ye, Li, and He]{guo2017deepfm}
H.~Guo, R.~Tang, Y.~Ye, Z.~Li, and X.~He.
\newblock Deepfm: a factorization-machine based neural network for ctr prediction.
\newblock \emph{arXiv preprint arXiv:1703.04247}, 2017.

\bibitem[Henighan et~al.(2020)Henighan, Kaplan, Katz, Chen, Hesse, Jackson, Jun, Brown, Dhariwal, Gray, et~al.]{henighan2020scaling}
T.~Henighan, J.~Kaplan, M.~Katz, M.~Chen, C.~Hesse, J.~Jackson, H.~Jun, T.~B. Brown, P.~Dhariwal, S.~Gray, et~al.
\newblock Scaling laws for autoregressive generative modeling.
\newblock \emph{arXiv preprint arXiv:2010.14701}, 2020.

\bibitem[Hoffmann et~al.(2022)Hoffmann, Borgeaud, Mensch, Buchatskaya, Cai, Rutherford, Casas, Hendricks, Welbl, Clark, et~al.]{hoffmann2022training}
J.~Hoffmann, S.~Borgeaud, A.~Mensch, E.~Buchatskaya, T.~Cai, E.~Rutherford, D.~d.~L. Casas, L.~A. Hendricks, J.~Welbl, A.~Clark, et~al.
\newblock Training compute-optimal large language models.
\newblock \emph{arXiv preprint arXiv:2203.15556}, 2022.

\bibitem[Hu et~al.(2024)Hu, Tu, Han, He, Cui, Long, Zheng, Fang, Huang, Zhao, et~al.]{hu2024minicpm}
S.~Hu, Y.~Tu, X.~Han, C.~He, G.~Cui, X.~Long, Z.~Zheng, Y.~Fang, Y.~Huang, W.~Zhao, et~al.
\newblock Minicpm: Unveiling the potential of small language models with scalable training strategies.
\newblock \emph{arXiv preprint arXiv:2404.06395}, 2024.

\bibitem[Jiang et~al.(2024)Jiang, Sablayrolles, Roux, Mensch, Savary, Bamford, Chaplot, Casas, Hanna, Bressand, et~al.]{jiang2024mixtral}
A.~Q. Jiang, A.~Sablayrolles, A.~Roux, A.~Mensch, B.~Savary, C.~Bamford, D.~S. Chaplot, D.~d.~l. Casas, E.~B. Hanna, F.~Bressand, et~al.
\newblock Mixtral of experts.
\newblock \emph{arXiv preprint arXiv:2401.04088}, 2024.

\bibitem[Kaplan et~al.(2020)Kaplan, McCandlish, Henighan, Brown, Chess, Child, Gray, Radford, Wu, and Amodei]{kaplan2020scaling}
J.~Kaplan, S.~McCandlish, T.~Henighan, T.~B. Brown, B.~Chess, R.~Child, S.~Gray, A.~Radford, J.~Wu, and D.~Amodei.
\newblock Scaling laws for neural language models.
\newblock \emph{arXiv preprint arXiv:2001.08361}, 2020.

\bibitem[Lee et~al.(2022)Lee, Kim, Kim, Cho, and Han]{lee2022autoregressive}
D.~Lee, C.~Kim, S.~Kim, M.~Cho, and W.-S. Han.
\newblock Autoregressive image generation using residual quantization.
\newblock In \emph{Proceedings of the IEEE/CVF Conference on Computer Vision and Pattern Recognition}, pages 11523--11532, 2022.

\bibitem[Li et~al.(2023)Li, Li, Savarese, and Hoi]{li2023blip}
J.~Li, D.~Li, S.~Savarese, and S.~Hoi.
\newblock Blip-2: Bootstrapping language-image pre-training with frozen image encoders and large language models.
\newblock In \emph{International conference on machine learning}, pages 19730--19742. PMLR, 2023.

\bibitem[Liu et~al.(2024)Liu, Feng, Xue, Wang, Wu, Lu, Zhao, Deng, Zhang, Ruan, et~al.]{liu2024deepseek}
A.~Liu, B.~Feng, B.~Xue, B.~Wang, B.~Wu, C.~Lu, C.~Zhao, C.~Deng, C.~Zhang, C.~Ruan, et~al.
\newblock Deepseek-v3 technical report.
\newblock \emph{arXiv preprint arXiv:2412.19437}, 2024.

\bibitem[Luo et~al.(2024)Luo, Cao, Sun, Yu, Huang, Yuan, Lin, Zheng, Wang, Hu, et~al.]{luo2024qarm}
X.~Luo, J.~Cao, T.~Sun, J.~Yu, R.~Huang, W.~Yuan, H.~Lin, Y.~Zheng, S.~Wang, Q.~Hu, et~al.
\newblock Qarm: Quantitative alignment multi-modal recommendation at kuaishou.
\newblock \emph{arXiv preprint arXiv:2411.11739}, 2024.

\bibitem[Ouyang et~al.(2022)Ouyang, Wu, Jiang, Almeida, Wainwright, Mishkin, Zhang, Agarwal, Slama, Ray, et~al.]{ouyang2022training}
L.~Ouyang, J.~Wu, X.~Jiang, D.~Almeida, C.~Wainwright, P.~Mishkin, C.~Zhang, S.~Agarwal, K.~Slama, A.~Ray, et~al.
\newblock Training language models to follow instructions with human feedback.
\newblock \emph{Advances in neural information processing systems}, 35:\penalty0 27730--27744, 2022.

\bibitem[Pi et~al.(2020)Pi, Zhou, Zhang, Wang, Ren, Fan, Zhu, and Gai]{pi2020search}
Q.~Pi, G.~Zhou, Y.~Zhang, Z.~Wang, L.~Ren, Y.~Fan, X.~Zhu, and K.~Gai.
\newblock Search-based user interest modeling with lifelong sequential behavior data for click-through rate prediction.
\newblock In \emph{Proceedings of the 29th ACM International Conference on Information \& Knowledge Management}, pages 2685--2692, 2020.

\bibitem[Rafailov et~al.(2023)Rafailov, Sharma, Mitchell, Manning, Ermon, and Finn]{rafailov2023direct}
R.~Rafailov, A.~Sharma, E.~Mitchell, C.~D. Manning, S.~Ermon, and C.~Finn.
\newblock Direct preference optimization: Your language model is secretly a reward model.
\newblock \emph{Advances in Neural Information Processing Systems}, 36:\penalty0 53728--53741, 2023.

\bibitem[Rajbhandari et~al.(2020)Rajbhandari, Rasley, Ruwase, and He]{rajbhandari2020zero}
S.~Rajbhandari, J.~Rasley, O.~Ruwase, and Y.~He.
\newblock Zero: Memory optimizations toward training trillion parameter models.
\newblock In \emph{SC20: International Conference for High Performance Computing, Networking, Storage and Analysis}, pages 1--16. IEEE, 2020.

\bibitem[Rajput et~al.(2024)Rajput, Mehta, Singh, Hulikal~Keshavan, Vu, Heldt, Hong, Tay, Tran, Samost, et~al.]{rajput2024recommender}
S.~Rajput, N.~Mehta, A.~Singh, R.~Hulikal~Keshavan, T.~Vu, L.~Heldt, L.~Hong, Y.~Tay, V.~Tran, J.~Samost, et~al.
\newblock Recommender systems with generative retrieval.
\newblock \emph{Advances in Neural Information Processing Systems}, 36, 2024.

\bibitem[Ren and Sutherland(2024)]{ren2024learning}
Y.~Ren and D.~J. Sutherland.
\newblock Learning dynamics of llm finetuning.
\newblock \emph{arXiv preprint arXiv:2407.10490}, 2024.

\bibitem[Rendle(2010)]{rendle2010factorization}
S.~Rendle.
\newblock Factorization machines.
\newblock In \emph{2010 IEEE International conference on data mining}, pages 995--1000. IEEE, 2010.

\bibitem[Ricci et~al.(2010)Ricci, Rokach, and Shapira]{ricci2010introduction}
F.~Ricci, L.~Rokach, and B.~Shapira.
\newblock Introduction to recommender systems handbook.
\newblock In \emph{Recommender systems handbook}, pages 1--35. Springer, 2010.

\bibitem[Shao et~al.(2024)Shao, Wang, Zhu, Xu, Song, Bi, Zhang, Zhang, Li, Wu, et~al.]{shao2024deepseekmath}
Z.~Shao, P.~Wang, Q.~Zhu, R.~Xu, J.~Song, X.~Bi, H.~Zhang, M.~Zhang, Y.~Li, Y.~Wu, et~al.
\newblock Deepseekmath: Pushing the limits of mathematical reasoning in open language models.
\newblock \emph{arXiv preprint arXiv:2402.03300}, 2024.

\bibitem[Shazeer(2020)]{shazeer2020glu}
N.~Shazeer.
\newblock Glu variants improve transformer.
\newblock \emph{arXiv preprint arXiv:2002.05202}, 2020.

\bibitem[Shoeybi et~al.(2019)Shoeybi, Patwary, Puri, LeGresley, Casper, and Catanzaro]{shoeybi2019megatron}
M.~Shoeybi, M.~Patwary, R.~Puri, P.~LeGresley, J.~Casper, and B.~Catanzaro.
\newblock Megatron-lm: Training multi-billion parameter language models using model parallelism.
\newblock \emph{arXiv preprint arXiv:1909.08053}, 2019.

\bibitem[Si et~al.(2024)Si, Guan, Sun, Zang, Lu, Hui, Cao, Yang, Zheng, Leng, et~al.]{si2024twin}
Z.~Si, L.~Guan, Z.~Sun, X.~Zang, J.~Lu, Y.~Hui, X.~Cao, Z.~Yang, Y.~Zheng, D.~Leng, et~al.
\newblock Twin v2: Scaling ultra-long user behavior sequence modeling for enhanced ctr prediction at kuaishou.
\newblock In \emph{Proceedings of the 33rd ACM International Conference on Information and Knowledge Management}, pages 4890--4897, 2024.

\bibitem[Thoppilan et~al.(2022)Thoppilan, De~Freitas, Hall, Shazeer, Kulshreshtha, Cheng, Jin, Bos, Baker, Du, et~al.]{thoppilan2022lamda}
R.~Thoppilan, D.~De~Freitas, J.~Hall, N.~Shazeer, A.~Kulshreshtha, H.-T. Cheng, A.~Jin, T.~Bos, L.~Baker, Y.~Du, et~al.
\newblock Lamda: Language models for dialog applications.
\newblock \emph{arXiv preprint arXiv:2201.08239}, 2022.

\bibitem[Wang et~al.(2024)Wang, Cao, Fu, Gai, and Zhou]{wang2024home}
X.~Wang, J.~Cao, Z.~Fu, K.~Gai, and G.~Zhou.
\newblock Home: Hierarchy of multi-gate experts for multi-task learning at kuaishou.
\newblock \emph{arXiv preprint arXiv:2408.05430}, 2024.

\bibitem[Yang et~al.(2020)Yang, Zhu, Zhang, Wang, and Yuan]{DBLP:journals/corr/abs-2010-05525}
X.~Yang, Y.~Zhu, Y.~Zhang, X.~Wang, and Q.~Yuan.
\newblock Large scale product graph construction for recommendation in e-commerce.
\newblock \emph{CoRR}, abs/2010.05525, 2020.

\bibitem[Zheng et~al.(2024)Zheng, Hou, Lu, Chen, Zhao, Chen, and Wen]{zheng2024adapting}
B.~Zheng, Y.~Hou, H.~Lu, Y.~Chen, W.~X. Zhao, M.~Chen, and J.-R. Wen.
\newblock Adapting large language models by integrating collaborative semantics for recommendation.
\newblock In \emph{2024 IEEE 40th International Conference on Data Engineering (ICDE)}, pages 1435--1448. IEEE, 2024.

\bibitem[Zhou et~al.(2018)Zhou, Zhu, Song, Fan, Zhu, Ma, Yan, Jin, Li, and Gai]{zhou2018deep}
G.~Zhou, X.~Zhu, C.~Song, Y.~Fan, H.~Zhu, X.~Ma, Y.~Yan, J.~Jin, H.~Li, and K.~Gai.
\newblock Deep interest network for click-through rate prediction.
\newblock In \emph{Proceedings of the 24th ACM SIGKDD international conference on knowledge discovery \& data mining}, pages 1059--1068, 2018.

\bibitem[Zhu et~al.(2024)Zhu, Wei, Lu, and Chen]{zhu2024scaling}
L.~Zhu, F.~Wei, Y.~Lu, and D.~Chen.
\newblock Scaling the codebook size of vqgan to 100,000 with a utilization rate of 99\%.
\newblock \emph{arXiv preprint arXiv:2406.11837}, 2024.

\bibitem[Ziegler et~al.(2019)Ziegler, Stiennon, Wu, Brown, Radford, Amodei, Christiano, and Irving]{ziegler2019fine}
D.~M. Ziegler, N.~Stiennon, J.~Wu, T.~B. Brown, A.~Radford, D.~Amodei, P.~Christiano, and G.~Irving.
\newblock Fine-tuning language models from human preferences.
\newblock \emph{arXiv preprint arXiv:1909.08593}, 2019.

\end{thebibliography}

\clearpage

\appendix
\section*{Appendix}
\section{Contributions}

Within each role, authors are listed alphabetically by their first name. 
Names marked with * denote individuals who have departed from our team. 

% \definecolor{damaiblue}{RGB}{0, 0, 100}
% \definecolor{damaigreen}{RGB}{0, 100, 0}
% \definecolor{damaired}{RGB}{100, 0, 0}
\begin{multicols}{2}
\noindent
\textbf{ Core Contributors} \\
 Guorui Zhou \\
 Jiaxin Deng \\
 Jinghao Zhang \\
 Kuo Cai \\
 Lejian Ren \\
 Qiang Luo \\
 Qianqian Wang \\
 Qigen Hu \\
 Rui Huang \\
 Shiyao Wang \\
 Weifeng Ding* \\
 Wuchao Li \\
 Xinchen Luo \\
 Xingmei Wang \\
 Zexuan Cheng \\
 Zixing Zhang

\noindent
\textbf{ Contributors} \\
 Bin Zhang\\
 Boxuan Wang \\
 Chaoyi Ma \\
 Chengru Song \\
 Chenhui Wang \\
 Di Wang \\
 Dongxue Meng\\
 Fan Yang \\
 Fangyu Zhang \\
 Feng Jiang \\
 Fuxing Zhang \\
 Gang Wang \\
 Guowang Zhang \\
 Han Li \\
 Hengrui Hu \\
 Hezheng Lin* \\
 Hongtao Cheng \\
 Hongyang Cao \\
 Huanjie Wang \\
 Jiaming Huang \\
 Jiapeng Chen \\
 Jiaqiang Liu \\
 Jinghui Jia \\
 Kun Gai\\
 Lantao Hu \\
 Liang Zeng \\
 Liao Yu \\
 Qiang Wang \\
 Qidong Zhou\\
 Shengzhe Wang \\
 Shihui He \\
 Shuang Yang \\
 Shujie Yang \\
 Sui Huang \\
 Tao Wu \\
 Tiantian He \\
 Tingting Gao\\
 Wei Yuan \\
 Xiao Liang \\
 Xiaoxiao Xu \\
 Xugang Liu \\
 Yan Wang\\
 Yi Wang \\
 Yiwu Liu \\
 Yue Song \\
 Yufei Zhang \\
 Yunfan Wu \\
 Yunfeng Zhao \\
 Zhanyu Liu
\end{multicols}

\newpage

\section{Implementation Details of Online A/B Test}
\label{appendix:online_ab}

In this section, we present the implementation details of OneRec in online A/B testing. In recommendation systems, a user's request typically triggers various system modules to generate real-time recommendation results. However, in practical applications, the massive QPS (the peak QPS can exceed 400k) necessitates substantial resources to handle such high concurrency. To address this, our system incorporates a caching mechanism: for each user request, the system returns $k$ recommendation results. Apart from the items actually exposed, the remaining items are stored as candidates in a cache pool. When the system experiences a high QPS load, cached results are retrieved for display, achieving a trade-off between resource usage and real-time performance. Thus, we broadly categorize QPS into real-time and degraded (cached) traffic, and OneRec’s online experiment specifically upgrades this degraded portion.  
There are two primary reasons for this experimental setup:

1. The previous caching mechanism significantly sacrificed the benefits of timeliness, affecting user experience during peak evening hours with high request volumes. While ``disabling the caching mechanism'' would incur substantial resource costs, OneRec’s highly efficient end-to-end pipeline and optimized MFU drastically reduce the system’s OPEX while delivering notable performance improvements.  

2. OneRec represents an entirely new architecture, introducing a fresh paradigm for technical iteration, business optimization, and team collaboration. We use this portion of traffic as a starting point to continuously explore technical boundaries and team collaboration mechanisms, building a robust foundation for handling more traffic.

\begin{figure}
    \centering
    \includegraphics[width=0.9\textwidth]{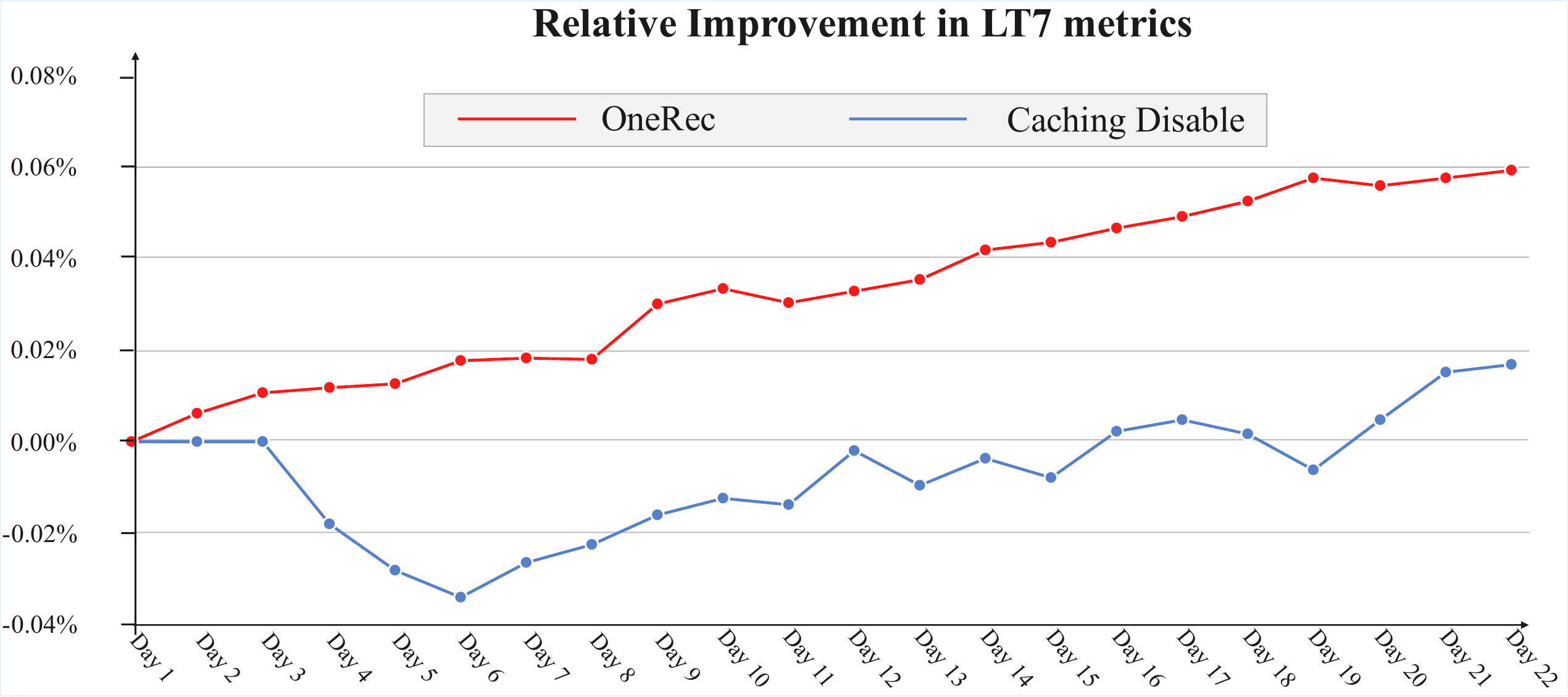}
    \caption{Comparative analysis of OneRec vs. Caching Disabled Architecture on LT7 growth trends.}
    \label{fig:lt7}
\end{figure}

As mentioned in Section \ref{online_ab_test}, our experimental group traffic is 5\%, with OneRec applied to 25\% of the degraded traffic within this group. Despite this limited scope, we observe significant performance gains across two scenarios, achieving 0.54\% and 1.24\% improvements in app stay time. For a more rigorous comparison, we allocate an additional 1\% experimental group with caching disabled (all traffic requesting real-time recommendations). Even against this baseline, OneRec demonstrates superior performance (shown in Table \ref{online_ab2}). We also observe the LT7 metric growth patterns between OneRec and the caching disabled strategy. Figure \ref{fig:lt7} indicates that OneRec exhibits significantly stronger improvement trends.

\begin{table}[H]
    \centering
    \caption{The absolute improvement of OneRec compared to the current multi-stage system and caching disabled experimental group (all traffic requesting real-time recommendations) in the online A/B testing setting.}
    \renewcommand{\arraystretch}{1.5} % 行高
    \setlength{\tabcolsep}{8pt}      % 列间距
    \begin{tabular}{p{3cm}|p{3.5cm}|p{3.5cm}|p{4cm}}
    \toprule
    Scenarios & Online Metrics &  vs. Current System &  vs. Caching Disabled\\
    \hline
    %\midrule
    \multirow{10}{*}{Kuaishou}&App Stay Time &+0.54\% &+0.20\%\\
    &LT7 & +0.05\%& +0.03\%\\
    &Watch Time & +1.98\%&+0.75\%\\
    &Video View &+2.52\% & +1.79\%\\
    &Engagement Depth & +1.78\%& +1.30\%\\
    &Like & +2.43\%&+0.88\%\\
    &Follow & +3.24\%&+1.29\%\\
    &Comment & +5.27\%& +3.18\%\\
    &Collect & +2.93\%&+0.73\%\\
    &Foward & +5.90\%&+4.92\%\\
    \hline
    \multirow{10}{*}{Kuaishou Lite}&App Stay Time & +1.24\% & +0.55\%\\
    &LT7 & +0.08\%&+0.02\%\\ 
    &Watch Time & +3.28\% & +1.58\%\\
    &Video View &+3.39\% & +1.71\%\\
    &Engagement Depth &+2.89\%& +2.49\%\\
    &Like &+1.49\% & -1.71\%\\
    &Follow &+2.28\% & +0.89\%\\
    &Comment &+3.20\% & +0.60\%\\
    &Collect &+1.91\% & -1.03\%\\
    &Foward &+3.48\% & +1.35\%\\
    \bottomrule
    \end{tabular}
    \label{online_ab2}
\end{table}

Through rigorous online A/B testing, our OneRec system has successfully replaced the original caching mechanism and now serves 25\% of the traffic in Kuaishou's main scenarios.

\section{Case Study for Tokenization}
\label{appendix:tokenizator}

\subsection{Representation Cases}

\begin{figure}
    \centering
    \includegraphics[width=0.85\textwidth]{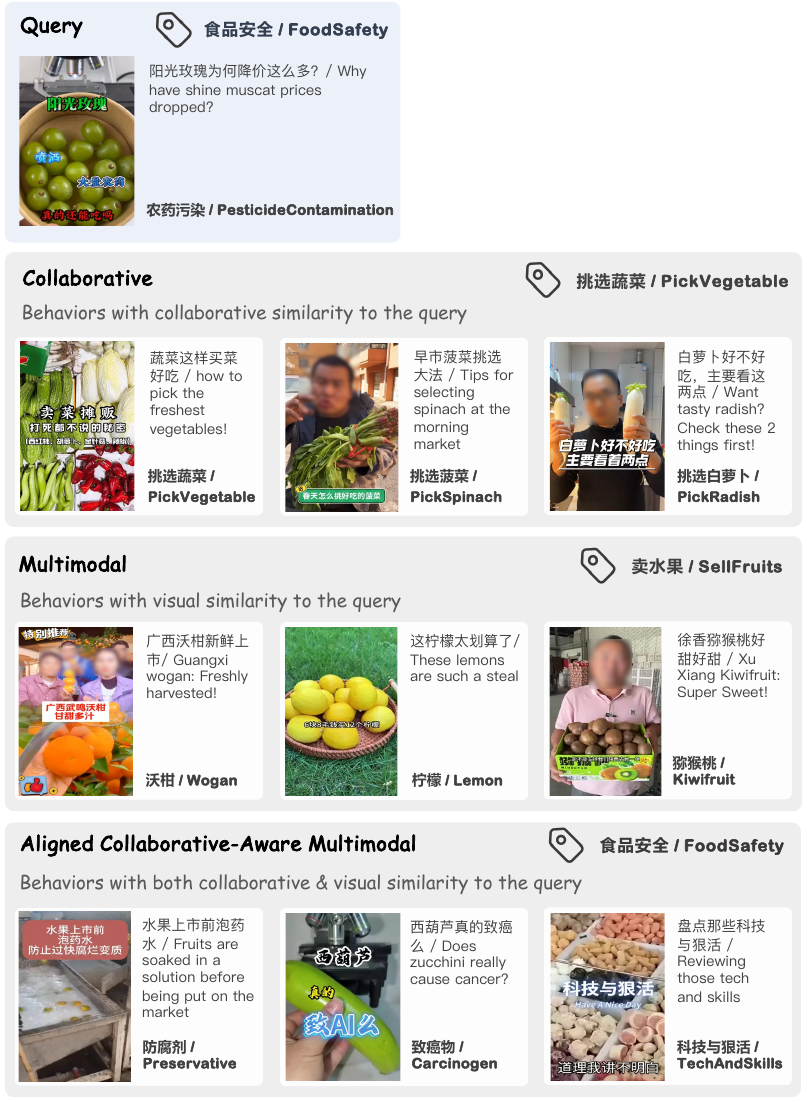}
    \caption{Cases of top-ranked videos retrieved from user history triggered by the query using different representation types.}
    \label{fig:repr1}
\end{figure}

\begin{figure}
    \centering
    \includegraphics[width=0.85\textwidth]{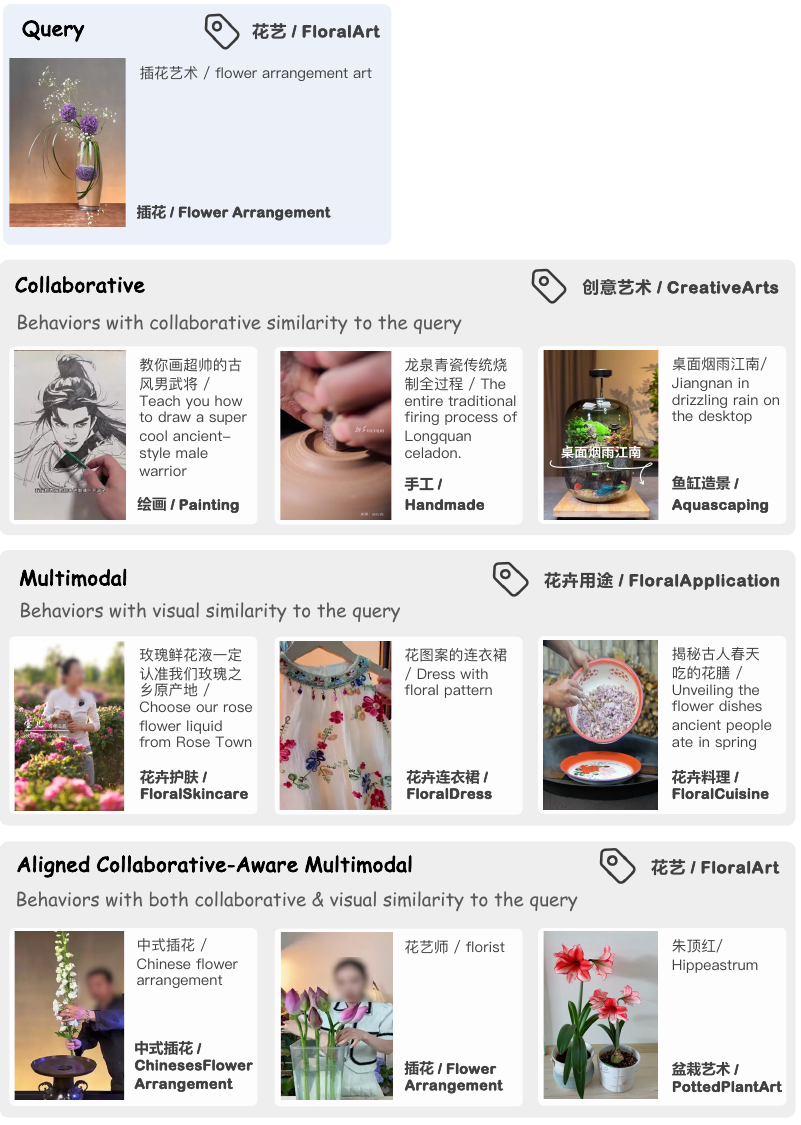}
    \caption{Cases of top-ranked videos retrieved from user history triggered by the query using different representation types.}
    \label{fig:repr2}
\end{figure}

\begin{figure}
    \centering
    \includegraphics[width=0.85\textwidth]{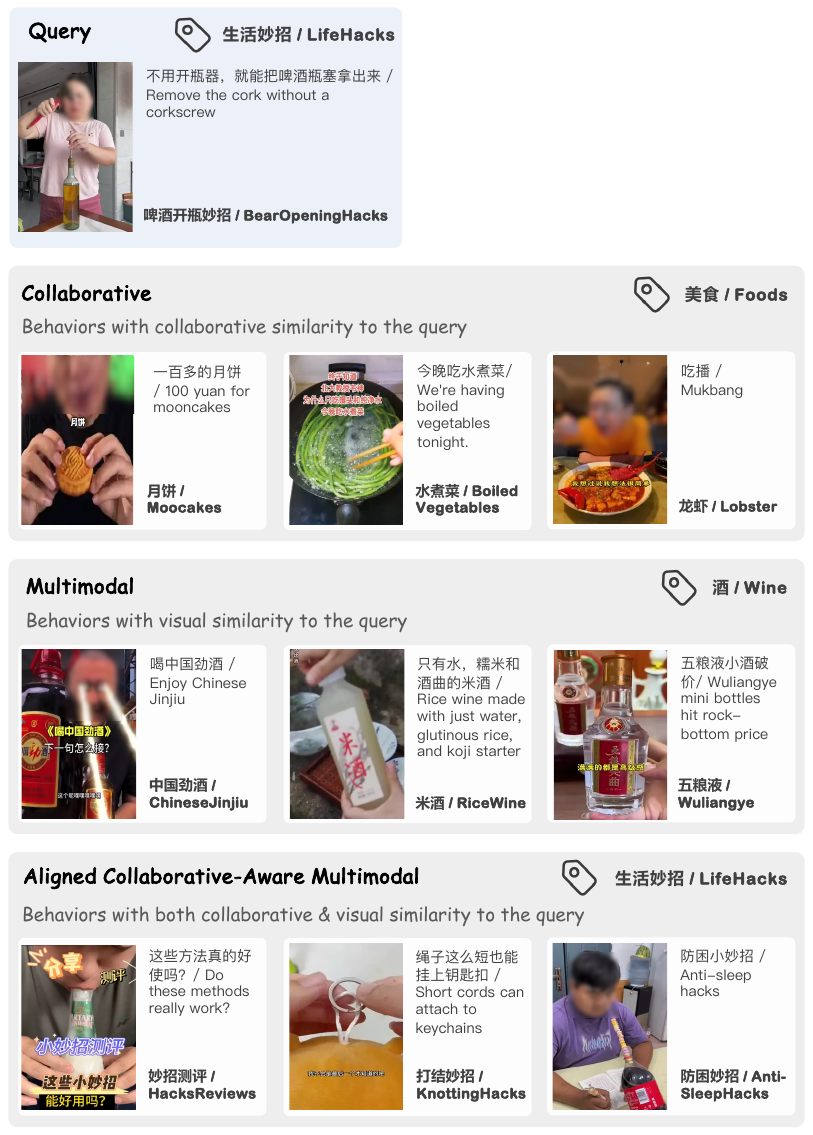}
    \caption{Cases of top-ranked videos retrieved from user history triggered by the query using different representation types.}
    \label{fig:repr3}
\end{figure}

To assess our aligned collaborative-aware multimodal representations, we contrast them with collaborative representations from traditional RS and pure multimodal representations extracted from caption/visual/OCR features. Figure~\ref{fig:repr1}, Figure~\ref{fig:repr2}, and Figure~\ref{fig:repr3} present illustrative cases demonstrating video retrieval results from user history for query videos when leveraging different representations.  

Our analysis reveals that collaborative representations—trained solely on collaborative signals—capture co-occurrence patterns but lack semantic relevance. This results in retrieved videos exhibiting categorical misalignment with query videos, as exemplified by painting content retrieved for a floral art query in Figure~\ref{fig:repr2} (row 2). Conversely, pure multimodal representations retrieve videos with surface-level feature similarities (e.g., shared visual elements like fruit in Figure~\ref{fig:repr1} (row 3) or wine in Figure~\ref{fig:repr3} (row 3)) yet fundamental categorical discrepancies relative to query videos. In contrast, our representations integrate multimodal and collaborative signals, enabling the retrieval of videos with multifaceted relevance. % that preserves both semantic coherence and feature-level similarities.
This demonstrates that our representations overcome the limitations of unimodal ones by jointly modeling content semantics and behavioral patterns.

\subsection{Tokenization Cases}

\begin{figure}[H]
    \centering
    \includegraphics[width=0.9\textwidth]{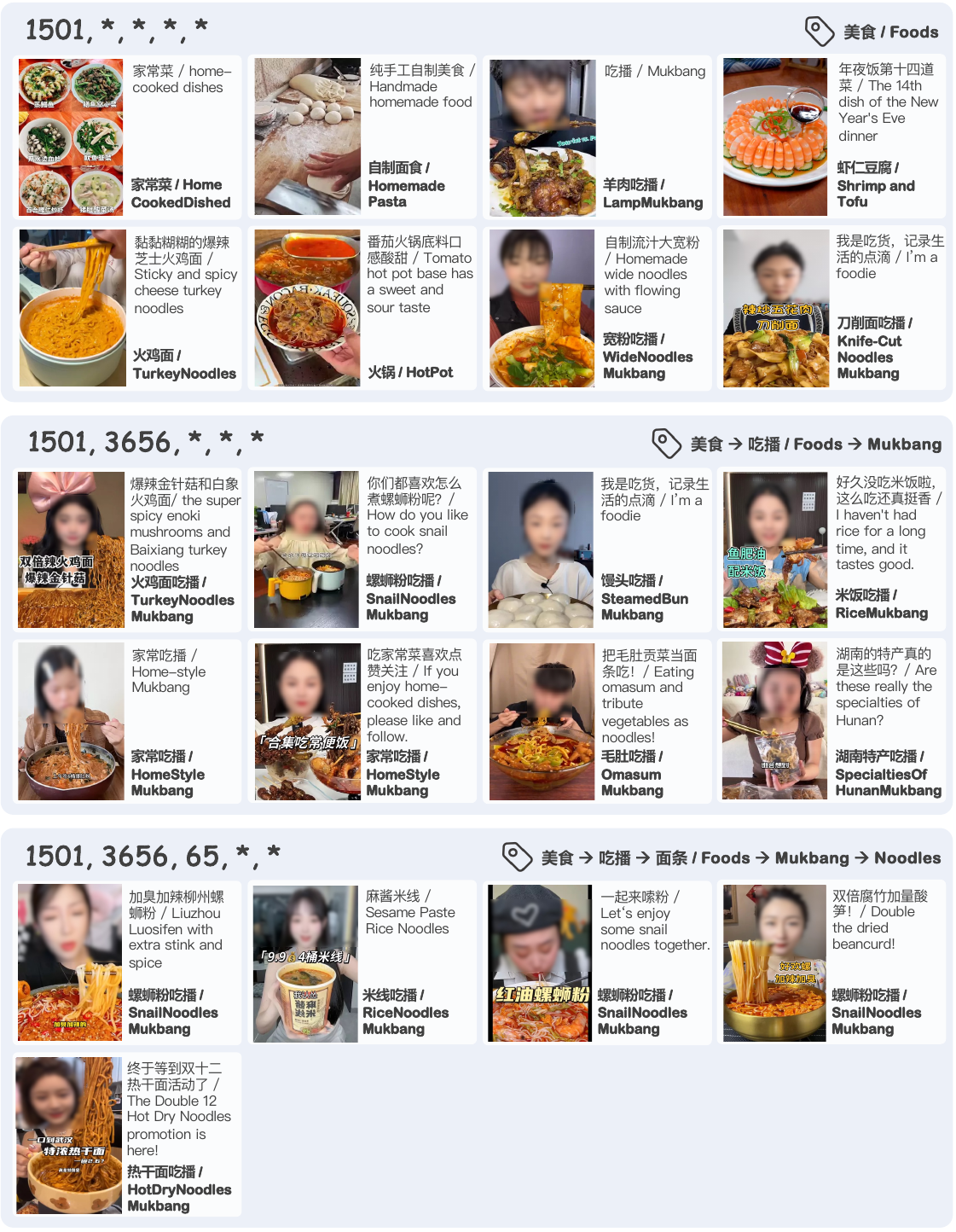}
    \caption{Cases of coarse-to-fine item semantic identifiers generated by RQ-Kmeans when $L_t=5$.}
    \label{fig:rqkmeans_1}
\end{figure}

\begin{figure}[H]
    \centering
    \includegraphics[width=0.9\textwidth]{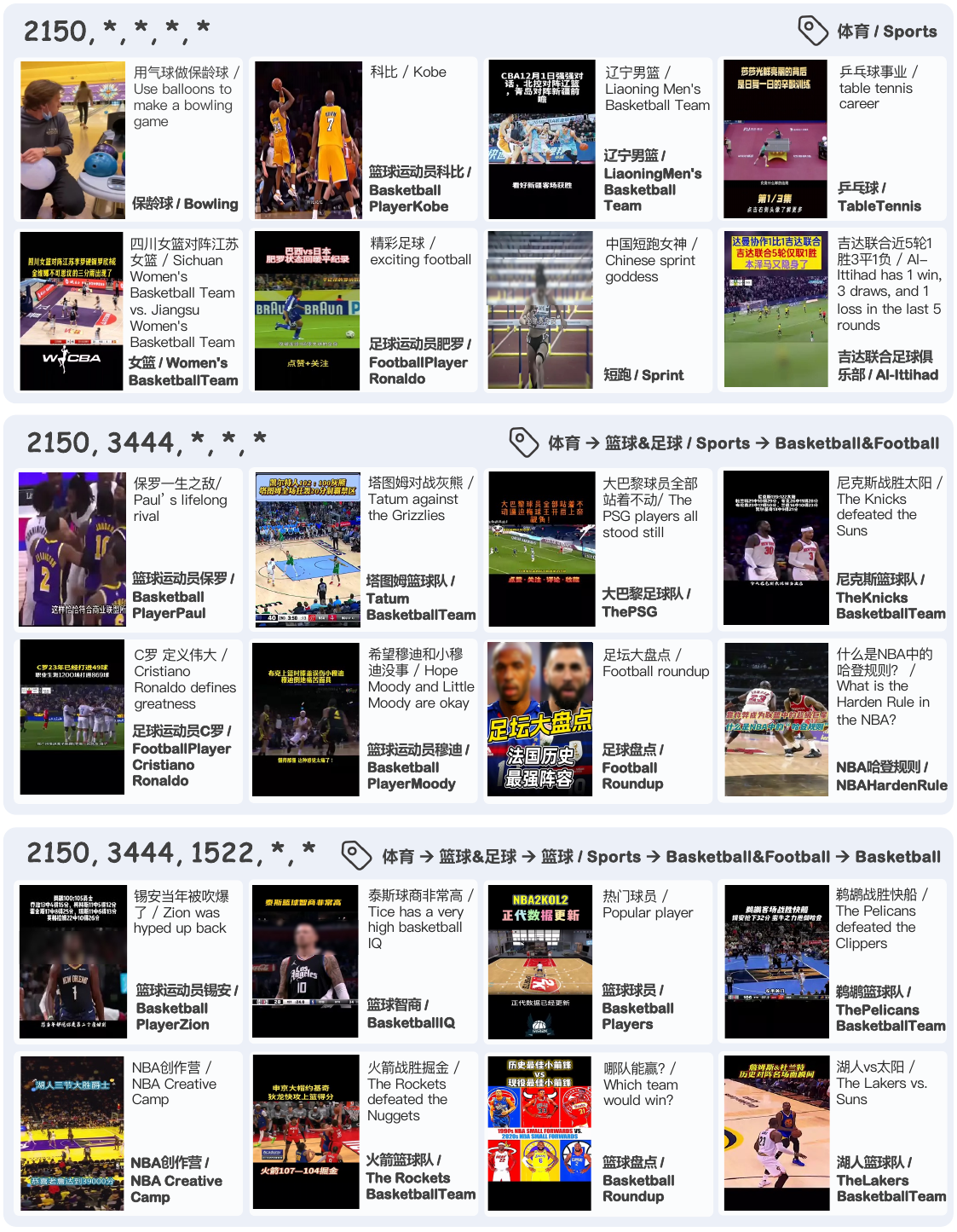}
    \caption{Cases of coarse-to-fine item semantic identifiers generated by RQ-Kmeans when $L_t=5$.}
    \label{fig:rqkmeans_2}
\end{figure}

We present cases of discrete item semantic identifiers generated by RQ-Kmeans in Figure~\ref{fig:rqkmeans_1} and Figure~\ref{fig:rqkmeans_2}. Our tokenization method can produce coarse-to-fine item semantic identifiers, where the first codeword indicates the coarsest category, and the categories of the second and third codewords become increasingly finer.

\section{Notations}
We summarize key notations used in this paper in Table~\ref{tab:notation_part1} and Table~\ref{tab:notation_part2}. 

\begin{table}[h]
\centering
\caption{Notation and Symbol Definitions in OneRec (Part 1)}
\begin{tabular}{|c|p{12cm}|}
\hline
\multicolumn{2}{|c|}{\textbf{General Notation}} \\
\hline
$d_{\text{model}}$ & Model hidden dimension (embedding dimension) \\
$L_t$ & Number of quantization layers in tokenization (set to 3) \\
$N_t$ & Codebook size for each quantization layer \\
$\{s^{1}_m, s^{2}_m, \ldots, s^{L_t}_m\}$ & Coarse-to-fine semantic identifiers for item $m$ \\
\hline
\multicolumn{2}{|c|}{\textbf{Item Tokenization}} \\
\hline
$d_{t}$ & Embedding dimension in tokenization (set to 512) \\
$N_M$ & The number of original multimodal token vectors of an item (set to 1280) \\
$\mathbf{M}$ & Multimodal token vectors from miniCPM-V-8B, $\mathbf{M} \in \mathbb{R}^{N_M \times d_{t}}$ \\
$N_{\tilde{M}}$ & The number of compressed multimodal token vectors of an item (set to 4) \\
$\mathbf{Q}^{(i)}$ & Query tokens in QFormer at layer $i$, $\mathbf{Q}^{(i)} \in \mathbb{R}^{N_{\tilde{M}} \times d_{t}}$ \\
$\tilde{\mathbf{M}}$ & Compressed multimodal representation after QFormer, $\tilde{\mathbf{M}} \in \mathbb{R}^{N_{\tilde{M}} \times d_{t}}$ \\
$N_{c}$ & Number of QFormer layers (set to 4) \\
$\mathcal{R}^{(l)}$ & Residual vectors at quantization layer $l$ \\
$\mathcal{C}^{(l)}$ & Codebook (K-means centroids) at quantization layer $l$ \\
$\bm{c}^{(l)}_k$ & $k$-th centroid in the codebook at layer $l$ \\
$s^{l}_i$ & Semantic identifier for item $i$ at quantization layer $l$ \\
$\mathcal{D}_{pair}$ & Dataset of item pairs with high collaborative similarity \\
$\tau$ & Temperature coefficient for item-to-item loss \\
$\text{sim}(\cdot, \cdot)$ & Similarity function used in item-to-item contrastive loss \\
$\mathcal{B}$ & A batch of $\mathcal{D}_{pair}$
\\
$t^k$ & The $k$-th caption token \\
\hline
\multicolumn{2}{|c|}{\textbf{Multi-Scale Feature Engineering}} \\
\hline
$L_s$ & Length of short-term behavior sequence (set to 20) \\
$L_p$ & Length of positive-feedback behavior sequence (set to 256) \\
$L_l$ & Length of lifelong behavior sequence (set to 2000) \\
$\mathbf{f}_u$ & Concatenated user static features before dense transformation \\
$\mathbf{f}_s$ & Concatenated short-term behavior features before dense transformation \\
$\mathbf{f}_p$ & Concatenated positive-feedback behavior features before dense transformation \\
$\mathbf{f}_l$ & Concatenated lifelong behavior features before dense transformation \\
$\mathbf{e}_{\text{*}}$ & Individual feature embeddings (e.g., $\mathbf{e}_{\text{uid}}, \mathbf{e}_{\text{gender}}, \mathbf{e}_{\text{age}}$ for user static) \\
$\mathbf{e}_{\text{*}}^s$ & Feature embeddings in the short-term pathway (e.g., $\mathbf{e}_{\text{vid}}^s, \mathbf{e}_{\text{aid}}^s, \mathbf{e}_{\text{tag}}^s$, etc.) \\
$\mathbf{e}_{\text{*}}^p$ & Feature embeddings in the positive-feedback pathway \\
$\mathbf{e}_{\text{*}}^l$ & Feature embeddings in the lifelong pathway \\
$\mathbf{h}_u$ & User static pathway representation, $\mathbf{h}_u \in \mathbb{R}^{1 \times d_{\text{model}}}$ \\
$\mathbf{h}_s$ & Short-term pathway representation, $\mathbf{h}_s \in \mathbb{R}^{L_s \times d_{\text{model}}}$ \\
$\mathbf{h}_p$ & Positive-feedback pathway representation, $\mathbf{h}_p \in \mathbb{R}^{L_p \times d_{\text{model}}}$ \\
$\mathbf{v}_l$ & Processed lifelong features before QFormer compression, $\mathbf{v}_l \in \mathbb{R}^{L_l \times d_{\text{model}}}$ \\
$\mathbf{h}_l^{(i)}$ & Query vectors at QFormer layer $i$ in the lifelong pathway \\
$\mathbf{h}_l$ & Final lifelong pathway representation, $\mathbf{h}_l \in \mathbb{R}^{N_q \times d_{\text{model}}}$ \\
$N_q$ & Number of query tokens in lifelong pathway compression (set to 128) \\
$N_l$ & Number of QFormer blocks in lifelong pathway (set to 2) \\
$M$ & Threshold for hierarchical clustering termination \\
\hline
\end{tabular}
\label{tab:notation_part1}
\end{table}

\begin{table}[h]
\centering
\caption{Notation and Symbol Definitions in OneRec (Part 2)}
\begin{tabular}{|c|p{12cm}|}
\hline
\multicolumn{2}{|c|}{\textbf{Encoder-Decoder Architecture}} \\
\hline
$L_{\text{enc}}$ & Number of transformer encoder layers \\
$L_{\text{dec}}$ & Number of transformer decoder layers \\
$\mathbf{e}_{\text{pos}}$ & Positional embeddings, $\mathbf{e}_{\text{pos}} \in \mathbb{R}^{(1 + L_s + L_p + N_q) \times d_{\text{model}}}$ \\
$\mathbf{z}^{(i)}$ & Hidden states at encoder layer $i$ \\
$\mathbf{z}_{\text{enc}}$ & Final encoder output \\
$\mathbf{d}^{(i)}_m$ & Decoder hidden states for item $m$ at layer $i$ \\
$\mathcal{S}_m$ & Input sequence for item $m$: $\{s_{[\mathrm{BOS}]}, s^1_m, s^2_m, \cdots, s^{L_t}_m\}$ \\
$s_{[\mathrm{BOS}]}$ & Beginning-of-sequence token \\
$N_{\text{experts}}$ & Number of expert networks in MoE layers \\
$k$ & Top-$k$ routing strategy parameter in MoE \\
$\text{Gate}_j(\mathbf{x})$ & Gating weights for $j$-th expert in MoE layer \\
$\text{Expert}_j(\mathbf{x})$ & Output of $j$-th expert network in MoE layer \\
\hline
\multicolumn{2}{|c|}{\textbf{Preference Alignment \& Reinforcement Learning}} \\
\hline
$\pi_\theta$ & Policy model with parameters $\theta$ \\
$\pi_{\theta_{old}}$ & Old policy model (before update) \\
$\pi_{\theta_{old}}'$ & Modified old policy with early clipping \\
$G$ & Number of generated samples per user \\
$K$ & Number of samples selected for format reward \\
$r_i$ & Reward for generated item $i$ (P-Score) \\
$A_i$ & Advantage for generated item $i$ \\
$\epsilon$ & Clipping parameter in ECPO \\
$\delta$ & Early clipping parameter in ECPO ($\delta > 0$) \\
$\mathcal{J}_{ECPO}(\theta)$ & ECPO optimization objective \\
$\text{sg}(\cdot)$ & Stop gradient operation \\
\hline
\multicolumn{2}{|c|}{\textbf{Industrial Constraints}} \\
\hline
$I_{\text{legal}}$ & Set of legal (valid) generated items \\
$I_{\text{viral}}$ & Set of viral content items \\
$f$ & Optimal proportion threshold for viral content \\
$\alpha$ & Down-weighting factor for viral content reward ($0 < \alpha < 1$) \\
\hline
\end{tabular}
\label{tab:notation_part2}
\end{table}

\end{document}